\documentclass[
 reprint,
 amsmath,amssymb,
 aps,
]{revtex4-1}

\usepackage{graphicx}
\usepackage{dcolumn}
\usepackage{bm}

\usepackage{color}
\newcommand{\sv}{\color{black}}
\newcommand{\svn}{\color{black}}

\begin{document}

\preprint{APS/123-QED}

\title{{\sv Predictability of large-scale atmospheric motions: Lyapunov exponents and error dynamics}}

\author{St\'ephane Vannitsem}
 \email{svn@meteo.be}
\affiliation{%
 Royal Meteorological Institute of Belgium\\
 Meteorological and Climatological Research \\
Avenue Circulaire, 3, 1180 Brussels \\
Belgium
}%

\bigskip
\bigskip
{\large Accepted in Chaos}
\bigskip
\bigskip

\date{\today}

\begin{abstract}
The deterministic equations describing the dynamics of the atmosphere (and of the climate system) are known to display the property of 
sensitivity 
to initial conditions. In the ergodic theory of chaos this property is usually quantified by computing the Lyapunov exponents. In this
review, these quantifiers computed in a hierarchy of atmospheric models (coupled or not to an ocean) are analyzed, {\sv together with
their local counterparts known as the local or finite-time Lyapunov exponents}. {\sv It is shown in particular that the variability of
the local Lyapunov exponents (corresponding to the dominant Lyapunov exponent) decreases when the model resolution increases}. The dynamics of 
(finite-amplitude) initial condition errors in these models {\sv is also reviewed}, and in general found to display a complicated growth 
far from the asymptotic estimates provided by the Lyapunov exponents. The implications of these results 
for operational (high resolution) atmospheric and climate modelling are also discussed.
\end{abstract}

\pacs{Valid PACS appear here}
\maketitle

{\bf
The models describing the dynamics of the atmosphere (and of the climate system) display 
the property of sensitivity to initial conditions, i.e. any small error introduced in the
initial conditions will grow in time until it reaches a level at which the forecast becomes useless. 
The atmosphere can therefore be considered as displaying a chaotic dynamics. In this review, the
dynamics of the initial condition errors is explored through their usual quantifiers 
known as the Lyapunov exponents (valid for infinitesimally small initial errors and 
infinite times), together with the analysis of the dynamics of finite-size errors, 
in a {\sv series} of {\sv atmospheric} models {\sv of increasing complexity}. 

}

\section{Introduction \label{intro}}

In the early fifties, the development of digital computers opened the possibility to 
perform weather forecasts using equations based on the
laws of hydrodynamics and thermodynamics. The first
successfull attempts have been obtained using a set of simplified equations based on two
approximations, the hydrostatic equilibrium and the approximate geostrophic balance, which respectively
postulate that the vertical pressure gradient force is equal to minus the gravitational force 
and the Coriolis force is approximately balanced by the horizontal pressure gradient force \cite{Charney1948}. 
These assumptions reduce considerably the number of pronostic equations and are at the
origin of the well-known quasi-geostrophic system of equations \cite{Holton1979, Haltiner1980}.
Since then important progresses have been made and up-to-date 
models based on primitive equations are used for forecasting purposes covering a large range of space scales and
vertical levels \cite{Kalnay2003}. These are 
supplemented by a considerable amount of physical parameterizations to simulate
cloud, rain and ice development, radiative transfers, surface interactions, and the impact of
sub-grid scale dynamics, among others.

The prospect of weather forecasting has rapidly raised the
question of the limits of predictability. A lot of effort
have then been devoted to answer this question whose basic properties were already 
identified in \cite{Thompson1957}. {\sv Thompson} \cite{Thompson1957} notably shows that the imperfect knowledge of the initial conditions induces a 
progressive degradation of weather forecasts. In other words 
 a small error committed on the initial conditions of the system will grow in time until it
reaches the size of the distance between two randomly chosen weather situations.
This property, presently known as the property of sensitivity to initial conditions (or initial states),
has been subsequently discovered in the numerical integration of a low-order deterministic system  by
\cite{Lorenz1963} -- based on the 
convection model originally developed by  Saltzman \cite{Saltzman1962} and presently known as the Lorenz model -- 
indicating the intrinsic nature of this dynamical property.
This pioneering work has raised a lot of interest in the community of atmospheric and climate sciences, and 
a lot of researches have been devoted to the analysis of 
sensitivity to initial conditions in a hierarchy of atmospheric and climate models, ranging from two-dimensional barotropic 
models \cite[e.g.][]{Basdevant1981, Legras1985, Lorenz1969} 
, quasi-geostrophic models \cite[e.g.][]{McWilliams1981, Vallis1983, Malguzzi1990, Barkmeijer1993, Schubert2015, Trevisan2001, Vannitsem1997, Vannitsem1998, Vannitsem2001, Morss2009}, 
global circulation models based on primitive equations \cite[e.g.][]{Schubert1989, Tribbia1988, Tribbia2004}, 
high-resolution atmospheric (mesoscale) models \cite[e.g.][]{Hohenegger2007, Bei2007, Uboldi2015}, 
operational weather forecasting models \cite[e.g.][]{Leith1978, Lorenz1982, Dalcher1987, Vandendool1990, Stroe1993, Savijarvi1995, Simmons1995, Buizza2010, Herrera2016} 
and climate models \cite[e.g.][]{Shukla1981,Roads1987, Palmer1988, Brankovic1990, Goswami1993, Palmer1996, Chen1997, Corti1998, Vandendool2006, Mumu2007, Duan2009, Delsole2014}. 
See also the reviews of \cite{Ehrendorfer1997, Kalnay2003, Yoden2007, Trevisan2011}. 
These different works have explored this property and the limit of predictability at different space and 
timescales and they all reach the same conclusion that sensitivity to initial conditions is a generic  
property of models describing the atmosphere and the climate system.  

It was also realized that other sources of errors
are degrading the forecasts, namely errors associated with the absence
of description of a set of processes, errors related to the parameterizations of subgrid-scale
processes, numerical errors, boundary condition errors and external forcing errors.     
The three first sources are usually referred as {\it model error}, while the boundary condition errors
and external forcing errors are considered separately. The impact of the presence of
these errors on the forecasts has also been investigated in atmospheric models of various complexities 
\cite[e.g.][]{Leith1978, Lorenz1982, Dalcher1987, Simmons1995, Tribbia2004, Vannitsem2006, Boisserie2014}, 
but it is only recently that a theory has been developed \cite{Orrell2001, Vannitsem2002, Nicolis2003, Nicolis2004, Judd2008, Nicolis2009}, 
revealing the polynomial nature of the short term error dynamics, contrasting with the exponential-like behavior of initial
condition errors. 

In parallel to these investigations, the interest of mathematicians and theoretical physicists for the problem of 
sensitivity to initial conditions
raised considerably and led to the development of the ergodic theory of chaos for deterministic dynamical systems 
\cite[e.g.][]{Ruelle1979, Eckmann1985, Ruelle1989}, 
and to the development of important quantifiers of the property of sensitivity to initial conditions, the Lyapunov exponents. 
One central result of this theory is that
in the double limit of infinitesimally small initial
errors and infinitely long times,
the distance between initially close trajectories increases (or decreases) in an exponential 
fashion with a rate, referred as
the (largest or dominant) Lyapunov exponent, which is an intrinsic property of the system's
attractor \cite{Ruelle1989, Ott1993, Smith2007, Sprott2010, Cencini2010}. Deterministic systems displaying a positive Lyapunov 
exponent, and therefore displaying the property of sensitivity to initial conditions, are referred to as {\it{chaotic systems}}.
  
The chaotic nature of atmospheric flows has been investigated using tools of ergodic theory in a hierarchy 
of atmospheric and/or oceanic models ranging from low-order 
\cite{Yoden1985, Nese1993, Yoden1993, Nicolis1995, Broer2011}, to more sophisticated, intermediate-order, models describing 
barotropic (2-dimensional) flows \cite{Legras1985} and quasi-geostrophic flows 
\cite{Malguzzi1990, Yano1992, Vannitsem1997, Szunyogh1997, Snyder2003, Lucarini2007, Schubert2015}.
All the results support the chaotic nature of the atmospheric models, and by extension of the atmosphere itself.

In the real world, one is usually dealing with the dynamics
of finite size errors during a finite time period. The double limit appearing in the definition
of the Lyapunov exponents cannot be attained and leads inevitably
to consider a behavior related with the local properties of the
attractor. Therefore in order to get information independent of the choice of the 
initial conditions,  it is necessary to adopt a probabilistic approach. 
This aspect has been extensively investigated in the past years 
and a systematic theory of error growth has been developed in the context of atmospheric sciences
\cite{Benzi1989, Nicolis1991, Nicolis1992, Trevisan1993, Yoden1993, Vannitsem1994, Nicolis1995, Trevisan1998, Boffetta1998, Yamane2001, 
Chu2002, Nicolis2009} -- and also in parallel in the context of turbulence \cite{Aurell1996, Bohr1998} and references therein. 
The key point of the approach is to incorporate information on the inhomogeneity of the dynamical properties of the  
solutions on the underlying attractor. 

In particular, it has been shown 
that this practical limitation is responsible for a complex non-exponential initial behavior
of the mean error for short times \cite{Nicolis1995}.  After this transient period, the error 
at all scales follows the dynamics of the dominant Lyapunov vector  associated with the dominant 
Lyapunov exponents (provided the error is still sufficiently small), and subsequently saturates when the nonlinearities 
are playing a dominant role.  Similar investigations have subsequently been performed in more complex
convection and atmospheric models by investigating the local properties of the Lyapunov vectors associated to each exponent 
\cite{Vannitsem1994, Szunyogh1997, Vannitsem1997, Snyder2003}, and the variability of their
local finite-time counterparts known as the singular vectors \cite{Molteni1993, Buizza1995, Ehrendorfer1995, Gelaro2002}.
  This stream of ideas led to the development of what is know nowadays as {\it ensemble forecasts} that are operational in
many weather centers around the world and which provide probabilistic information on the evolution of the
atmosphere, as discussed in several reviews on ensemble forecasts \cite{Ehrendorfer1997, Kalnay2003, Yoden2007}.  

The Lyapunov exponents (and the variability {\sv of their finite-time counterparts} along the attractor of the system) are 
therefore key quantities for the  understanding of the predictability of the atmosphere (and of climate). 
In the present paper,  the computation of the Lyapunov exponents {\sv and of the statistical properties 
of the finite-time (or local) Lyapunov exponents }
in a hierarchy of atmospheric (and climate) models, is reviewed. Their relevance for the description of the
predictability in highly detailed atmospheric and climate models is then discussed. 

Section II is devoted to a general overview of the classical
deterministic modelling of the atmospheric dynamics (together with a very brief introduction of the large-scale upper 
ocean dynamics) and Section III to the description of
the computation of the Lyapunov exponents. Results obtained with a hierarchy of low-order to intermediate order
atmospheric (and climate) models are presented and discussed in Section IV. The dynamics of the error is then described in Section V.
Section VI is devoted to the future challenges in characterizing the predictability of atmospheric and climate flows.     

\section{Modeling the atmospheric and ocean dynamics \label{model}}
 
Traditional atmospheric (and climate) models are based
on the classical set of conservation laws of hydrodynamics \cite{Haltiner1980, Ghil1987}.  For the atmosphere, 
these include mass balance, moisture balance, momentum balance and energy balance.
These equations are complemented by a number of diagnostic relations such as the equation of state. 
The typical set of equations used for describing the dynamics of the atmosphere are,

the conservation of momentum, 
\begin{equation}
\frac{d{\vec{v}}}{dt} = - 2 \vec{\Omega} \times \vec{v} - \nabla \Phi - \frac{1}{\rho} \vec{\nabla} p - \frac{1}{\rho} \vec{\nabla} \cdot \vec{\vec{\sigma}}
\end{equation}
where $\vec{v}$, $\rho$, $p$, $\vec{\Omega}$, $\Phi$ {\sv and $\vec{\vec{\sigma}}$ } are the three-dimensional velocity field, the atmospheric density, the pressure, the angular
velocity of the Earth, the geopotential {\sv and the stress tensor}, respectively;

the conservation of mass,
\begin{equation}
\frac{1}{ \rho } \frac{d \rho }{dt} = - \vec{\nabla} \cdot \vec{v}   
\end{equation}
{\sv where $\rho$ is the density of air masses};

the ideal gas law,
\begin{equation}
p=\rho R T
\end{equation}
where $T$ is the temperature and $R$, the gas constant;

the thermodynamic equation,
\begin{equation}
c_p \frac{d}{d t} T - \frac{1}{\rho} \frac{d}{d t} p = Q 
\end{equation}
{\sv where $Q$ is the rate of heat per unit mass added to the fluid and $c_p$, the specific heat at constant pressure;}

and the conservation equation for the water vapor content, $q$, 
\begin{equation}
\frac{d q }{dt} = E-C   
\end{equation}
where $E$ and $C$ are the evaporation rate and the condensation rate, respectively.
This set of equations are often known as {\it the primitive equations} \cite{Houghton1984, Kalnay2003}.
They are further complemented by appropriate boundary conditions and with complicated radiative forcings and heat
exchanges, all contained in the term $Q$, and known as {\it diabatic processes}. In realistic numerical weather prediction 
$Q$, $E$ and $C$ play a crucial role, and should be complemented by physical packages describing the formation of clouds, the
development of rain, the chemical reactions, etc..., and their interaction with the dynamics described above.  

These equations are usually mapped to a spherical geometry at global scale or on a regional domain of interest, 
and often simplified by assuming the vertical scale of the motion to be small compared with the horizontal one.  The
equations are further reduced to a set of ordinary differential equations through spatial discretization  
using finite difference schemes or truncation of the infinite expansion of the field in an
appropriate functional basis, or both \cite{Haltiner1980}. 

Starting from this set of equations, the process
of forecasting consists first in identifying the phase space point that represents most adequately the
initial condition available from observation based on {\it data assimilation} techniques \cite{Daley1991}. 
The next step is to compute
numerically, by additional discretization in time, the trajectory of the dynamical system in phase
space, {\sv also known as numerical model integration}. To reach a high spatial resolution one includes the maximum number
of degrees of freedom compatible with the computing power available.  
Usually the complication of the structure of operational atmospheric and climate models precludes reliable 
statistical analysis or a systematic exploration of the behavior in terms of the parameters.

An important class of models of the atmospheric circulation which have been used
extensively for forecasting purposes, is provided by the quasi-geostrophic models
\cite{Holton1979, Vallis2006}. These models are obtained by adopting a number of assumptions in the full set of
balance equations, the most important of which are:
(i) the atmosphere is in hydrostatic equilibrium;
(ii) the wind and pressure fields are in approximate geostrophic equilibrium so that the
horizontal advection {\sv is essentially described by} the non-divergent velocity field;
(iii) the dynamical equations contain only the dominant contributions of a Taylor
expansion of the Coriolis force.

More formally, these approximations are justified through the natural scaling of the dominant large scale flows in both the
atmosphere and the ocean at mid-latitudes, for which the pressure gradient is in approximate balance with the Coriolis force. The predominance
of this approximate balance is associated with a non-dimensional number known as the Rossby number, $Ro=U/(fL)$, where $U$ and $L$ are   
the typical horizontal velocity and length scales of the large scale flows and the coriolis parameter $ f=2 \Omega \sin(\phi)$
where $\Omega$ is the amplitude of the angular velocity of the Earth {\sv and $\phi$ the latitude}. For the
atmosphere {\sv at midlatitudes} this number is of the order of $0.1$ and for the ocean of the order of $0.01$, see \cite{Vallis2006} for a more detailed
discussion on these scalings.
 
These simplifications led to an equation of conservation for the potential vorticity in pressure coordinates, 
\begin{equation}
q=\nabla^2 \psi + f  + {f_0}^2 \frac{\partial}{\partial p} \sigma^{-1} \frac{\partial \psi}{\partial p}
\end{equation}
where $\psi$ is the streamfunction, $f_0$, the dominant contribution of the
Coriolis force estimated at $\phi_0=45^\circ$, and $\sigma$, the static stability parameter,
\begin{equation}
\frac{\partial q}{\partial t} + (\vec{v} \cdot \vec{\nabla}) q =  F 
\label{vorticitepot1}
\end{equation}
where $\vec{v}=(-\partial \psi/\partial y, \partial \psi/\partial x)$ is the non-divergent horizontal velocity field and 
$F$ contains all the dissipative and forcing terms. This conservation law and the notion of potential vorticity has been
considerably exploited for the understanding of the large scale atmospheric dynamics, see e.g. \cite{Hoskins1985}.  
Note that in this setting the temperature in the atmosphere is given by 
\begin{equation}
T_a= - \frac{f_0 p}{R} \left ( \frac{\partial \psi}{\partial p} \right )_p 
\label{temp}
\end{equation}
where $R$ is the ideal gaz constant. 

Based on similar approximations, one can also deduce a conservation equation (\ref{vorticitepot1}) 
for the large scale dynamics of the upper layer of an ocean (considered as homogeneous) in which the potential 
vorticity $q$ is now 
\begin{equation}
q=\nabla^2 \Psi + \beta y  - \frac{1}{L^2_d} \Psi
\label{oceandyn}
\end{equation}
where $\beta=df/dy$ at $\phi_0=45^\circ$ and $L_d=\sqrt{g'H}/f_0$ with $H$ the depth of the water layer, and $g'=g(\rho'-\rho)/\rho$, 
the {\sv reduced acceleration of gravity where $\rho$ and $\rho'$ are the densities of two superimposed ocean layers, the lower one being an infinitely deep layer at rest} 
\cite{Vallis2006}. Note also that an important forcing of the ocean dynamics (present in the term $F$   
in the right hand side of \ref{vorticitepot1}) is the wind stress at the ocean surface expressed as,
\begin{equation}
\frac{\mathrm{curl}_z \vec{\tau}}{\rho H} = \frac{C}{\rho H} \nabla^2 (\psi_{lower}-\Psi_{upper}). 
\label{eq:stress}
\end{equation}
{\sv 
where the wind stress is proportional to the relative velocity between the wind in the lower 
atmospheric layer, $\vec{v}_{lower}$ and the flow in the ocean upper layer, $\vec{v}_{o,upper}$, namely $\vec{\tau}=C (\vec{v}_{lower}-\vec{v}_{o,upper})$. } 
The drag coefficient,  $C$, characterizes the strength of the mechanical coupling between the ocean and 
the atmosphere and is a  key bifurcation parameter in the coupled model that will be discussed later.

The dynamical {\sv systems} analysis presented below consists first to embed the
evolution of the system just described above in a space spanned by the ensemble of relevant variables, known
as the phase space. Typical phase space variables are the values of the meteorological fields at
grid points, or the coefficients of their expansions in an appropriate functional basis.
Their number is usually very high ($10^8$ or so in operational forecasting), unless drastic 
truncations leading to low-order or intermediate-order models are performed. 
Let us now focus on the specific models that will be used in the analyses that will follow. 

\subsection{Low-order models \label{low-order}}

Low-order models have flourished in various fields of science \cite{Sprott2010}, and in particular in atmospheric and
climate sciences
\cite[e.g.][]{Veronis1963, Lorenz1965, Charney1979, Charney1980, Yao1980, Reinhold1982, Lorenz1984, Cehelsky1987, Vallis1988, Maas1994, Jiang1995, Yoden1997, Vanveen2003, Dijkstra2005, Simonnet2005, Pierini2011, Vannitsem2014a, Vannitsem2014c, 
Decruz2016, Chen2016}. These simplified models containing key ingredients of the physics of the atmosphere and/or the ocean allow for clarifying important 
features of the underlying dynamical properties in phase space. 

For the present illustrative purpose, we will make use of a model of the atmospheric dynamics at midlatitudes developed by
Charney and Straus \cite{Charney1980}, referred as the CS model in the following,  and a recent extension of this model 
developed for the understanding of the coupled ocean-atmosphere dynamics \cite{Vannitsem2015a, Vannitsem2015b}. 
The latter is first presented in some details and the simplifications leading to the CS model will be outlined. 

The atmospheric model is based on the vorticity equation (\ref{vorticitepot1}) defined at two superimposed 
atmospheric levels, say 1 and 2,
 which constitutes a minimal representation for the development of the so-called baroclinic instabilities at the origin of
the main variability of the weather at midlatitudes \cite{Holton1979}. 
The ocean dynamics confined to a single homogeneous layer is based on a 
vorticity equation with the potential vorticity given by (\ref{oceandyn}). Finally an advection equation for the temperature
in the ocean considered as a passive scalar is incorporated in the model, 
\begin{equation}
\frac{\partial{T_o}}{\partial t} + (\vec{v}_o \cdot \vec{\nabla}) T_o = -\lambda (T_o-T_a)+E_R(t) 
\end{equation}
where $\vec{v}_o$ is the (non-divergent) ocean velocity, $E_R(t)$ the radiative input in the ocean and $-\lambda (T_o-T_a)$ the heat
exchange between the ocean and the atmosphere. For more details on the equations of the model and the parameters, 
see \cite{Vannitsem2015a, Vannitsem2015b}.     

The model is forced by short-wave radiations coming from the Sun and an energy balance scheme is redistributing the 
energy through long-wave radiation emissions and heat tranfer between the two components of the system. The energy
entering into the ocean is   
\begin{equation}\label{eq:fluxes_oc}
E_R (t) = -\sigma_B T_o^4 + \epsilon_a \sigma_B T_a^4 + R_o(t).
\end{equation}
where $\epsilon_a$ is the emissivity of the atmosphere, $\sigma_B$, the Stefan-Boltzman constant and $R_o(t)$ the net radiative 
input entering the ocean coming from the Sun. While for the atmosphere the radiative input is, 
\begin{equation}\label{eq:fluxes_atm}
E_{a,R}(t) = \epsilon_a \sigma_B T_o^4 - 2 \epsilon_a \sigma_B T_a^4 + R_a(t).
\end{equation}
where $R_a(t)$ is fixed to $R_o(t)/3$.  It is assumed that the temperature fields can be linearized around a reference 
temperature in both the atmosphere and the ocean as
\begin{subequations}\label{eq:lin}
\begin{align}
& T_a = T_{a,0} + \delta T_a, \label{lin_atm} \\
& T_o = T_{o,0} + \delta T_o, \label{lin_oc}
\end{align}
\end{subequations}
where $T_{a,0}$ and $T_{o,0}$ are spatially uniform temperatures. {\sv It is also assumed that the atmosphere is dry and is not affected by effects
associated with the development of rain, ice and clouds}.

Also let
\begin{subequations}\label{eq:rad}
\begin{align}
& R_a (t) = R_{a,0} (t) + \delta R_a(t) , \label{rad_atm} \\
& R_o (t) = R_{o,0} (t) + \delta R_o(t) , \label{rad_oc}
\end{align}
\end{subequations}
with $R_{o,0}(t) $ and  $R_{a,0}(t) $ are time dependent spatially uniform shortwave radiative forcings, and $\delta R_a (t)$ and $\delta R_o (t)$, the spatially
varying counterparts. 

\begin{figure*}
\includegraphics{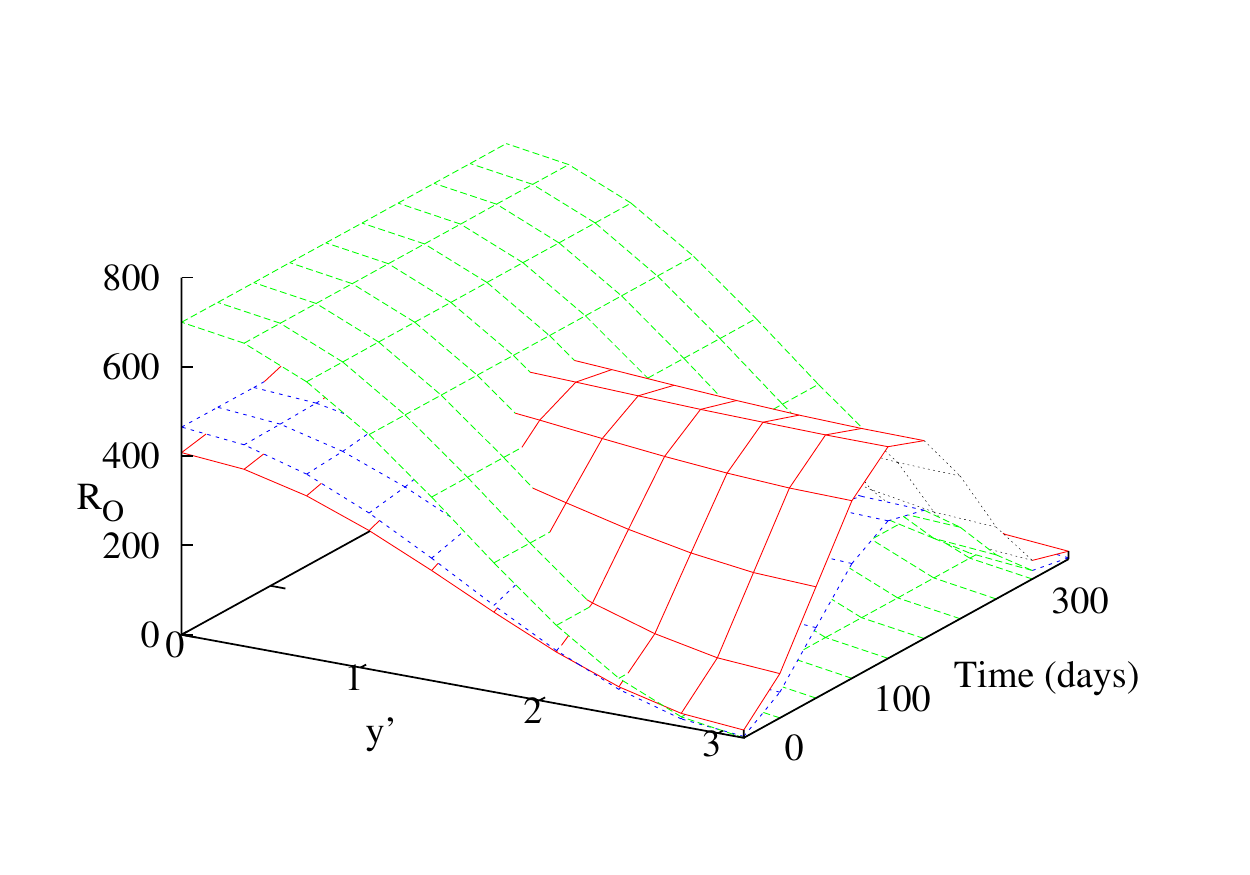}
\caption{
\label{insolation}
Analytical expression (\ref{radiation}) of the radiative input as a function of latitude and time for different values of $\kappa$, $\kappa=1$ (green),
$\kappa=0.5$ {\sv (blue)}, and $\kappa=0.3$, (red),{\sv  with $S_o=310$ W m$^{-2}$}.
The latitude is displayed in adimensional units on the domain $[0,\pi]$.  The unit along the vertical axis is W m$^{-2}$.
}
\end{figure*}

In order to mimick as close as possible the radiative input coming from the sun at midlatitudes, $R_{o} (t)$, used in the low-order model, is approximated as 
\begin{eqnarray}
R_o= R_{o,0}  & + & \delta R_o  =  S_o (1+\alpha \sin (\omega (t - \zeta))) \nonumber \\ 
& + & \kappa S_o \cos(y') (1 - 2 \alpha sin (\omega (t - \zeta)))  \nonumber \\
\label{radiation}
\end{eqnarray}
where $\omega=2 \pi /365$ days$^{-1}$, $\zeta=80$ days, and $y'$ is the latitude in non-dimensional units varying from $[0,\pi]$.
$\kappa$ is a free parameter {\sv varying between $]0,1]$} and $\alpha$ is fixed such that
the radiative input is never negative in the whole domain, $\alpha=\min(((1/\kappa)-1)/((1/\kappa)+2),0.5)$. This choice also implies that
the energy input in the non-autonomous case is reaching $0$ at $y=\pi$ at $t \approx 355$ days (Winter solstice).
Two free parameters are present in this relation: $S_o$, the energy input, and $\kappa$, the latitudinal contribution.
Figure \ref{insolation} displays $R_o$ for different values of $\kappa$, the smaller the value of $\kappa$, the larger the
seasonal variations. For $\kappa$=0.3, the seasonal variation is very similar to the actual evolution as discussed in \cite{Vannitsem2015a}.

A second important parameter largely influenced by the seasonal variations of the radiative input is the depth of the
upper ocean layer, known as the mixed layer, interacting directly with the atmosphere as discussed in \cite{Vannitsem2015a}.
In this context we choose the following relation for the ocean depth,
\begin{equation}
H (t) = D_{ref} \ln \left ( 1+\left ( \frac{500}{R_{o,0} (t)}\right )^3 \right) 
\label{H}
\end{equation}
where $D_{ref}$ is fixed to $100$ m in most of the integrations performed below, unless it is explicitly stated.

The atmospheric and oceanic fields are expanded in
Fourier series over the domain,
$(0 \le x' \le 2\pi/n, 0 \le y' \le \pi)$, where
$n$ is the aspect ratio between the
meridional and the zonal extents of the domain,
$n = 2 L_y/L_x $, and $x'=x/L$ and $y'=y/L$.

One retains the following set of modes for the dynamics within the ocean,
\begin{eqnarray}\label{eq:ocean_set}
\phi_1 & = &  2 sin(n x'/2) \sin(y'), \nonumber \\ 
\phi_2 & = &  2 \sin(n x'/2) \sin(2y'), \nonumber \\ 
\phi_3 & = &  2 \sin(n x') \sin(y'), \nonumber \\ 
\phi_4 & = &  2 \sin(n x') \sin(2y'), \nonumber \\ 
\phi_5 & = &  2 \sin(n x'/2) \sin(3y'), \nonumber \\ 
\phi_6 & = &  2 \sin(n x'/2) \sin(4y'), \nonumber \\ 
\phi_7 & = &  2  \sin(n x') \sin(3y'), \nonumber \\ 
\phi_8 & = &  2 \sin(n x') \sin(4y').
\end{eqnarray}
For temperature, the same set is used, except the modes $\phi_1$ and $\phi_5$ for which the spatial average is different from 0.
This allows to interpret the reference temperature within the ocean as a spatial temperature average.

For the atmosphere, we keep the same set of modes as
in \cite{Reinhold1982},

\begin{eqnarray}
F_1 & = &   \sqrt{2} \cos(y'),  \nonumber \\
F_2 & = &   2 \cos(n x') \sin(y'), \nonumber \\ 
F_3 & = &  2  \sin(n x') \sin(y'), \nonumber \\ 
F_4 & = &  \sqrt{2} \cos(2y'), \nonumber \\
F_5 & = &  2  \cos(n x') \sin(2y'),  \nonumber \\ 
F_6 & = &   2 \sin(n x') \sin(2y'), \nonumber \\ 
F_7 & = &  2 \cos(2 n x') \sin(y'), \nonumber \\ 
F_8 & = &  2  \sin(2 n x') \sin(y'), \nonumber \\ 
F_9 & = &  2  \cos(2 n x') \sin(2y'), \nonumber \\ 
F_{10} & = & 2 \sin(2 n x') \sin(2y'),   
\label{Fmodes}
\end{eqnarray}

All equations are then projected onto these sets of modes after linearizing the temperature equations around reference spatially averaged
temperatures. The projection is performed using the usual scalar product,
\begin{equation}
<f,g>= \frac{n}{2 \pi^2} \int_0^\pi dy' \int_0^{2 \pi/n} dx' f(x',y') g(x',y') 
\label{inneratmos}
\end{equation}
for the non-dimensional equations.
It leads to 8 ordinary differential equations (ODEs) for the dynamics within the ocean, one equation for the spatially averaged ocean temperature and 6 equations
for the anomaly temperature field within the ocean. In addition 20 ODEs are obtained for the atmosphere
, 10 for the barotropic streamfunction field
 $(\psi_1+\psi_2)/2$ and 10 for the baroclinic streamfunction field $\theta=(\psi_1-\psi_2)/2$ 
(also often referred to temperature due to its direct link with Eq. (\ref{temp}) in this setting). 
An additional equation for the spatially averaged atmospheric temperature is also deduced.
It forms a set of 36 ODEs which is fully described in the Supplement of \cite{Vannitsem2015a}.

The CS model \cite{Charney1980} is a simplified atmospheric version of the model just described above without ocean, 
for which the 
radiative input is directly introduced as a forcing of the baroclinic streamfunction equation {\sv in the form of}  a Newtonian relaxation 
toward an equilibrium baroclinic streamfunction field, $\theta^*$.  
The equilibrium solution, $\theta^*$, is chosen as
\begin{equation}
\theta^* = \theta^*_1 F_1
\end{equation}   
in adimensional units and constant in time. 
This model version is supposed to mimick the dominant dynamics of the atmosphere over a 
land surface with an idealized orography, which is given as  
\begin{equation}
h = h_2 F_2
\end{equation}   
also in non-dimensional units. The development of the fields in Fourier modes is limited to the first 6 modes of \ref{Fmodes},
leading to 12 ODEs. A full description of the model is given in \cite{Charney1980}.

The model equations are integrated in time using a second order Heun method.

\subsection{Intermediate order models \label{QG3}}

Intermediate order models have been developed in order to alleviate the limitations of high-resolution climate
models whose computer time demand is very high, but still providing a realistic dynamics of the processes of interest. 
These are typically truncated model versions with an horizontal resolution of a few hundred of kilometers for the atmosphere.   

Such a global model based on equation (\ref{vorticitepot1}) (and denoted as QG21L3 in the following) involving three levels 
along the vertical has been proposed in \cite{Marshall1993}. 
Thanks to the relative manageability of this model (1449 variables), an extensive analysis can be performed.

The model describes the evolution of the potential vorticity (\ref{vorticitepot1}) at three vertical levels, 200hPa,
500 hPa and 800 hPa \cite{Marshall1993}.  
The horizontal fields $Z$ are expanded in series of spherical harmonics $Y_{m,n}$ truncated triangularly at wavenumber 21:

\begin{equation}
Z(\lambda, \phi, t) = \sum_{n=0}^{21} \sum _{m=-n}^{m=n} Z_{m,n}(t) Y_{m,n} (\lambda, \phi)
\end{equation}
where $\lambda$, $\phi$, $m$ and $n$ are the longitude, the latitude, the zonal and total wavenumbers,
respectively. The index $n$ represents a total (two-dimensional) wavenumber on the sphere and
characterizes the size of the two-dimensional horizontal structures. The prognostic equation at
each level, $i$, can then be written in terms of the streamfunction $\psi$ and the potential vorticity 
$q$ as
\begin{equation}
\frac{\partial q_i}{\partial t} = - J(\psi_i, q_i) - D(\psi_i) + S_i
\label{vorticity}
\end{equation}
where $J(\psi,q)$ is the nonlinear jacobian operator, 
$\partial \psi/ \partial x \, \partial q /\partial y -\partial q / \partial x \, \partial \psi /\partial y  $. 
The linear term $D$ accounts for the
effects of Newtonian relaxation of temperature, a scale selective horizontal diffusion of vorticity
and temperature and a drag on the wind at the lower level whose coefficient depends on the
properties of the underlying surface. Finally the time-independent spatially varying source term,
$S_i$, constrains the solution of the model to an averaged, statistically stable, observed winter
climatology ("perpetual winter" conditions). Note that all the fields are computed in
non-dimensionalized units: the length unit is the earth radius (6371 km) and the time unit is half the
inverse of the angular velocity of the earth ($7.292 10^{-5} s^{-1}$). 
The model equations are integrated in time using a leapfrog scheme (together with a Robert-Asselin filter) 
with a time step of 1 hour starting from a realistic potential vorticity field. 
The model is fully described in Appendix A of \cite{Marshall1993}.

\section{The Lyapunov exponents \label{theory}}

The instantaneous fields of these models are represented by points in phase space and
as time elapses the phase space trajectories followed by the system's solutions tend to an invariant manifold,
to which one refers as the attractor. This reflects the dissipative character of meteorological and
climate phenomena. As we are interested in characterizing the instability of the flows generated by these models, let us 
focus on the dynamics of (infinitely) small amplitude errors and the computation of the Lyapunov exponents.

The evolution laws of a dynamical system like the ones presented in Section \ref{model} can be written in the synthetic form
\begin{equation}
\frac{d\vec{x}}{dt} = {\vec f}(\vec x, \lambda)
\label{equat}
\end{equation}
where $\vec x$ is a vector containing the entire set of relevant variables 
$\vec x$ = $( x_1, ..., x_n)$
such as temperature, wind velocity, ..., projected on the relevant set of modes (or grid points) as discussed in Section \ref{model}.
The {\sv functions} $\vec f$ represent the
effect of dynamical processes responsible for the change of $\vec x$, and 
$\lambda$ denotes a set of parameters such as emission or absorption coefficients, turbulent viscosity,  
etc. 

As mentioned in the Introduction, the initial state is never known exactly since the process of measurement and data
assimilation is always subjected to finite precision.
To clarify the implications of the presence of such an error we consider an initial state displaced slightly
from $\vec x (t_0)=\vec x_0$ by an initial error $\delta \vec x_0$.
This perturbed initial state generates a new trajectory in phase space  
and we define the instantaneous error vector as the vector joining the
points of the reference trajectory and the perturbed one at 
a given time, $\delta \vec x (t)$. Provided that this perturbation is sufficiently small, its dynamics
can be described by the linearized equation,
\begin{equation}
\frac{d\delta \vec x}{dt} =  \frac{\partial \vec f}{\partial \vec x}_{\vert \vec x(t)}
\delta \vec x
\label {linear}
\end{equation}
and a formal solution can be written as,
\begin{equation}
\delta \vec x (t) = {\bf M}(t,\vec{x}(t_0)) \delta \vec x (t_0)
\end{equation}
where the matrix $\bf{M}$, referred as the resolvent matrix, plays an important role in error growth dynamics 
as revealed when writing the Euclidean norm of the error,
\begin{eqnarray}
E_t & = & \vert \delta \vec x (t) \vert^2 = {\delta \vec x (t) }^T {\delta \vec x (t) } \nonumber \\ 
 & = & {\delta \vec x (t_0) }^T {\bf M}(t,\vec{x}(t_0))^T {\bf M}(t,\vec{x}(t_0)) {\delta \vec x (t_0) } \nonumber \\
\label{eucl}
\end{eqnarray}

One immediately realizes that the growth of $E_t$ is conditioned by the eigenvalues of the matrix
$\bf{M}^T \bf{M}$, where $(.)^T$ indicates transposition (and complex conjugation in complex space if necessary). 

In ergodic theory of chaotic systems, the double
limit of infinitely small initial errors and infinitely long times, is usually considered \cite[e.g.][]{Eckmann1985}. In
this limit the divergence of initially closed states is determined by the logarithm of the eigenvalues of the 
matrix $(\bf{M}^T \bf{M})^{2(t-t_0)}$ that are referred as the Lyapunov exponents.  
The full set of Lyapunov exponents of a system is called the Lyapunov spectrum which are usually represented in 
decreasing order. 
In the limit of $t \rightarrow \infty$, the eigenvectors of matrix $\bf{M}^T \bf{M}$, which are local properties of the 
flow and depend on the initial time $t_0$, are called the Forward Lyapunov vectors \cite{Legras1995}.

Notice that the eigenvalues of the matrix $\bf{S}=(\bf{M}^T \bf{M})^{2(t-t_0)}$ obtained for $t \rightarrow \infty$  
are
equivalent to the ones of the matrix $\bf{S'}=(\bf{M}\bf{M}^T)^{2(t-t_0)}$ when $t_0 \rightarrow -\infty$. 
On the contrary,
the eigenvectors of these {\sv two matrices $\bf{S}$ and $\bf{S'}$ -- denoted as $\vec{l}^+_i$ and $\vec{l}^-_i$, respectively -- } 
are not equivalent due to the asymmetric character of the
resolvent $\bf{M}$. The eigenvectors of $\bf{S'}$ are called the Backward Lyapunov vectors. 

Several techniques have been developed to numerically evaluate these Lyapunov exponents \cite{Kuptsov2012}. 
One of the most popular method consists in following the evolution of a set of
orthonormal vectors $\vec s_i$ chosen initially at random in the tangent space of the trajectory 
$\vec x(t)$.
This basis is regularly orthonormalized using the standard Gram-Schmidt method to avoid the
alignment of all the vectors along the unstable direction associated to the largest Lyapunov
exponent. After a rapid transient, the first vector of this set, free of any constraint, will tend
to the direction of maximal stretching associated to the largest Lyapunov exponent; the second
vector, orthogonal to the previous one, will tend to the second most unstable direction; and
so on. This set of orthonormal vectors evolving in the tangent space 
correspond asymptotically to the Backward Lyapunov vectors. 

These vectors and their properties were extensively discussed in recent years in the literature {\sv \cite[e.g.][]{Legras1995,Szunyogh1997, Vannitsem1997,Gelaro2002}}, 
in particular with respect to the significance of the eigenvectors of the matrices $S$ and $S'$. Note that
these vectors are not perturbations that are covariant with the dynamics
of the error in the tangent (linearized) space of the phase space trajectory. Other subspaces were then
introduced, $W_i$   
\begin{equation}
W_i(\vec x(t)) = \vec l^-_1 \oplus ... \oplus \vec l^-_i \cap \vec l^+_i \oplus ... \oplus \vec l^+_N
\label{wi}
\end{equation}
where $\oplus$ is the direct product \cite{Ruelle1979, Kuptsov2012, Schubert2015, Vannitsem2016}. Any vector
in this {\sv new} subspace is covariant with the (linearized) dynamics as,
\begin{equation}
{\bf M} (\tau ,\vec x(t')) \vec g_i(\vec x(t')) = \sigma_i (\tau, \vec x(t')) \vec g_i (\vec x(\tau)) 
\label{ampli}
\end{equation}
where $\sigma_i (\tau, \vec x(t'))$ is the amplification factor{\sv, and $\tau > t'$}. Note first that the basis $\{ \vec g_i\}$ do
not form an orthogonal basis and also that in the long time
limit, the amplifications give also access to the Lyapunov exponents,
\begin{equation}
\sigma_i =  \lim_{(\tau-t') \rightarrow \infty} \frac{1}{\tau-t'} ln \,\, (\sigma_i (\tau, \vec x(t')))
\end{equation}
The vectors $\{ \vec g_i\}$ are called the Covariant Lyapunov vectors.

The three approaches based on the Forward, Backward or Covariant Lyapunov vectors give the same Lyapunov spectrum.
However higher order properties like
the variance of the local amplification rates are not equal whether the Forward, Backward or Covariant Lyapunov
vectors are used, see \cite{Vannitsem2016}, except for the first Backward and last Forward 
Lyapunov vectors that have identical statistical properties as the first and last Covariant Lyapunov vectors, 
respectively.  

Since we will focus in the present review on the Lyapunov spectra and the properties of the dominant Lyapunov
exponent and vector, we will not discuss further the properties of the Covariant vectors and will leave the
interested reader to explore the recent literature on that subject \cite{Ginelli2007, Pazo2008, Kuptsov2012, 
Froyland2013, Schubert2015, Vannitsem2016}. We will however illustrate what is the variability (inhomogeneity) of
the {\sv instability properties on the }attractors of the different models by investigating the amplification rates, 
{\sv $\alpha_1 (\tau,t)= 1/(\tau-t) \ln (\sigma_1 (\tau, \vec x(t)))$,} 
along the dominant Backward (or Covariant) Lyapunov vector in the spirit 
of \cite{Gallez1991, Abarbanel1991, Nicolis1995, Vannitsem1997, Vannitsem2016}.   

\section{Lyapunov instabilities of atmospheric flows \label{lyap}}

This section is devoted to the description of the Lyapunov properties of chaotic solutions found so far in the hierarchy
of models discussed in Section \ref{model}. The purpose is to illustrate the modifications of these properties when
the number of variables of a model is increased and when dealing with a multi-scale system, and to highlight the open
questions arising nowadays in atmospheric and climate sciences concerning the problem of predictability. 

\subsection{Lyapunov exponents of the low-order atmospheric system \label{lyap_CSmodel}}

Let us focus on the Lyapunov properties of the 12-variable low-order system introduced in Section \ref{low-order}, 
the atmosphere CS model.

\begin{figure*}
\includegraphics{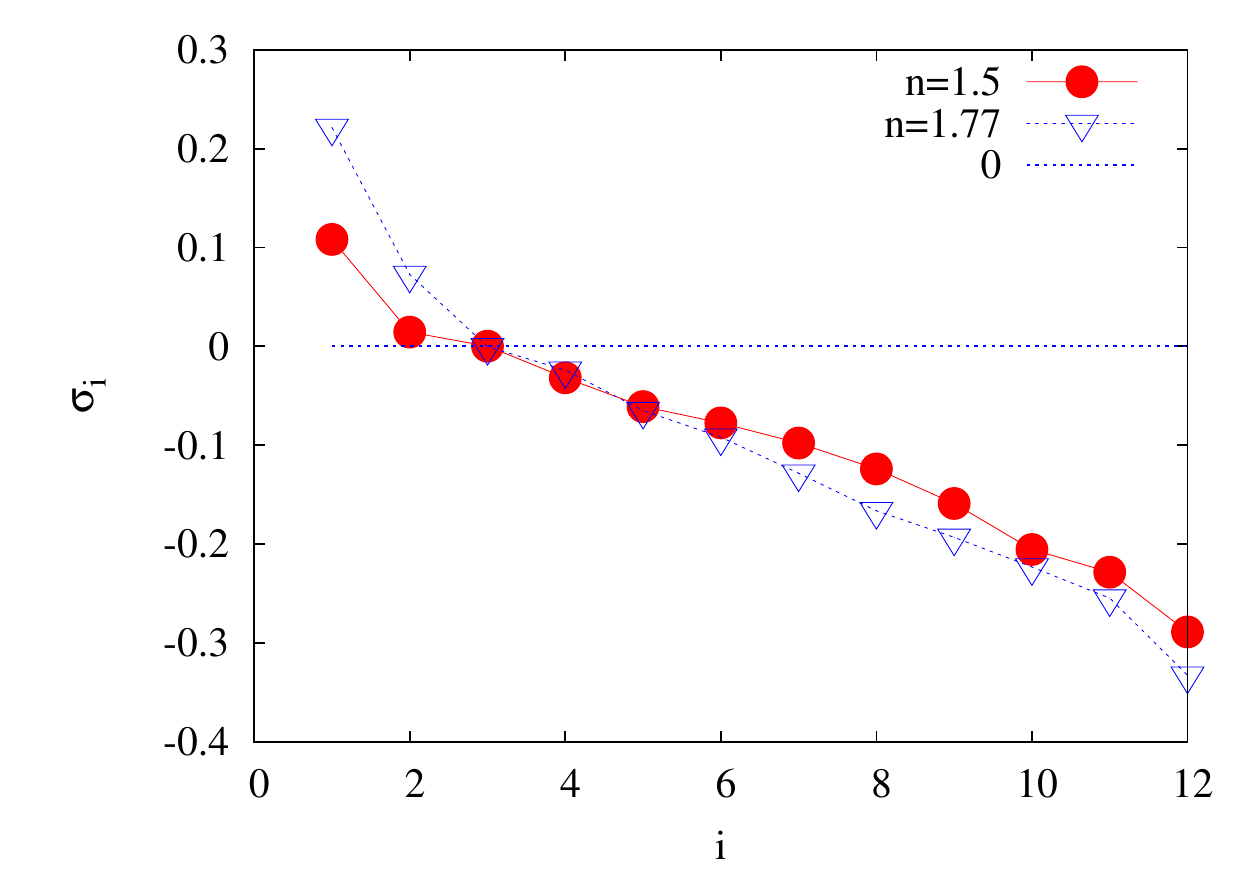}
\caption{
\label{CSspectra}
Lyapunov spectra of the CS model as obtained after 10,000 days of integrations for two different aspect ratios, 
$n=1.5$ (red filled circles) and $n=1.77$ (blue triangles). The values of the Lyapunov exponents are given in day$^{-1}$.}
\end{figure*}

Figure \ref{CSspectra} displays the Lyapunov spectra as obtained after an integration of 10,000 days for 
parameter values $\theta_1^*=0.20$, $h_2=0.1$ {\sv and for 2 different values of $n$}. 
Note that the other parameters defined in the original paper of Charney and Straus \cite{Charney1980} will not be discussed 
here for conciseness and are fixed to $2 k=k'=h"=0.0114$, $\sigma_0=0.2$, $L=5000/\pi$ km.

A clear picture emerges with 2 positive exponents, one 0, and 9 negatives ones for the two different 
aspect ratios explored, $n=1.5$ and $1.77$, the latter being originally used in \cite{Charney1980}. 
The solution is {\sv(hyper)}chaotic and displays a Lyapunov spectrum typical of low-order systems
such as the ones studied in the atmospheric context \cite{Malguzzi1990, Nicolis1995}, or
in a more general physical context \cite{Sprott2010}. Note that the dominant Lyapunov exponent for $n=1.77$ is of the same 
order as the amplitude of the dominant Lyapunov exponent found in more sophisticated atmospheric models as discussed later 
in Section \ref{QG3_lyap}.    

Interestingly, the spectrum 
is highly sensitive to the aspect ratio -- i.e. a smaller domain size in the zonal 
direction corresponds to a larger value of the aspect ratio -- indicating that instability 
properties of the flow depends crucially on the typical wavelengths present in the dynamical 
system. This is reminiscent of the sensitivity of the classical baroclinic instabilities as a function of the 
dominant wavelength of the perturbation \cite{Holton1979}.   

\begin{figure*}
\includegraphics{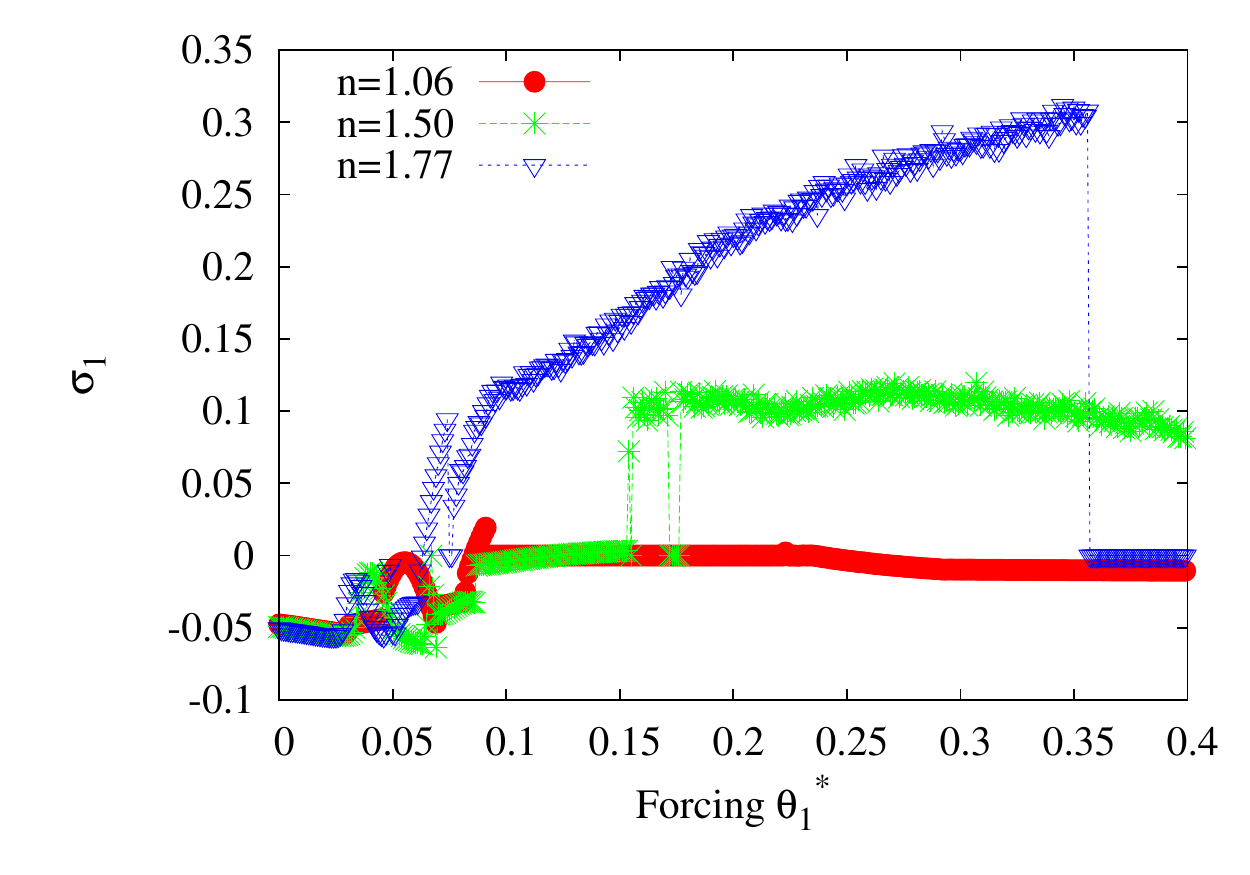}
\caption{
\label{CS1st}
First Lyapunov exponent of the CS model as a function of the thermal forcing $\theta_1^*$ for three different aspect ratios, 
$n=1.06$ (red filled circles), $n=1.5$ (green stars) and $n=1.77$ (blue triangles). The values of the Lyapunov exponents are given in day$^{-1}$.}
\end{figure*}

This specific dependence on $n$ is also visible when the key parameter $\theta_1^*$ associated to the
meridional variations of the radiative input in the domain is varied as illustrated in 
Fig. \ref{CS1st}, with a higher instability for the parameter $n=1.77$ corresponding to a smaller domain
size in the zonal direction. Windows of periodic solutions are also visible reflecting the complicate structure of the
bifurcation diagram for this model as usually found in other low-order models, \cite[e.g.][]{Ott1993}.       

\begin{figure*}
\includegraphics{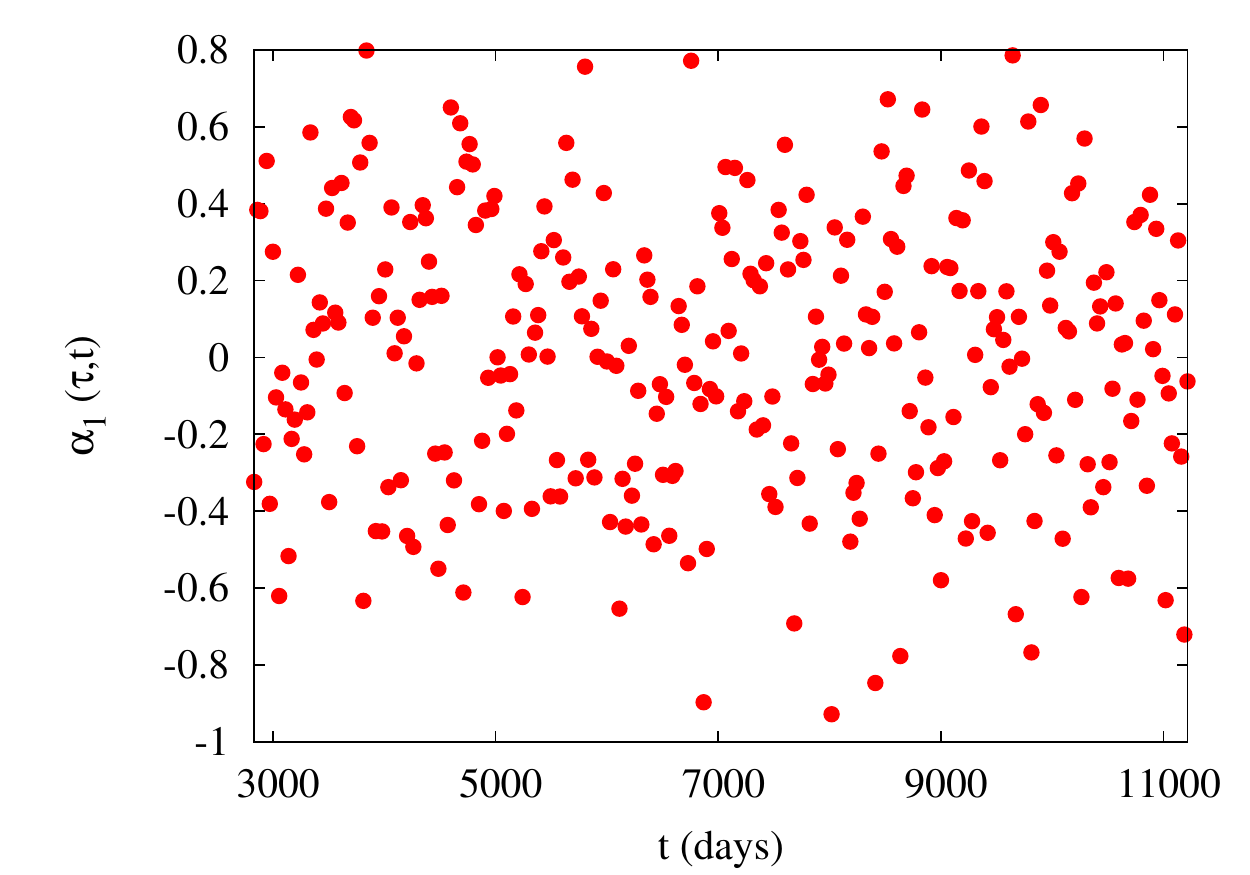}
\caption{
\label{CS1stloc}
Temporal variability of the local amplification rate $\alpha_1(\tau,t)$ for the CS model with $n=1.77$, $\theta_1^*=0.2$, 
and $\tau-t=0.0056$ days. The values are sampled every $28.4$ days.}
\end{figure*}

Another important aspect of the instability properties of this system is {\sv the high variability of the local Lyapunov exponents} on the
attractor. This is illustrated in Fig. \ref{CS1stloc} where the local amplification rate $\alpha_1(\tau,t)$
is plotted as a function of $t$ with $\tau-t=0.0056$ days and sampled every 28.4 days 
(corresponding to 250 non-dimensional time units). A very large variability of this amplification is visible
covering values from $[-1,1]$ day$^{-1}$. The standard deviation of this series is equal to $0.41$ day$^{-1}$, a value twice as large as the value of
the first Lyapunov exponent itself. 

This {\sv variability on the attractor of the system} indicates that the predictability of
atmospheric flows highly dependis on the specific underlying atmospheric situation. This natural variability of the 
weather skill is also experienced in real forecasts, see e.g. \cite{Branstator1986, Yoden2007}, but  
to a lesser extent than in the current model. This point will be taken up further in the next section while investigating a higher
resolution atmospheric model. 

\subsection{Lyapunov exponents of The QG3T21 model \label{QG3_lyap}}

Let us now turn to a more sophisticated atmospheric model described in Section \ref{QG3}. This model
has 1449 degrees of freedom \cite{Vannitsem1997}, and its solution is thus embedded in a phase space of fairly high
dimension. 

\begin{figure*}
\includegraphics{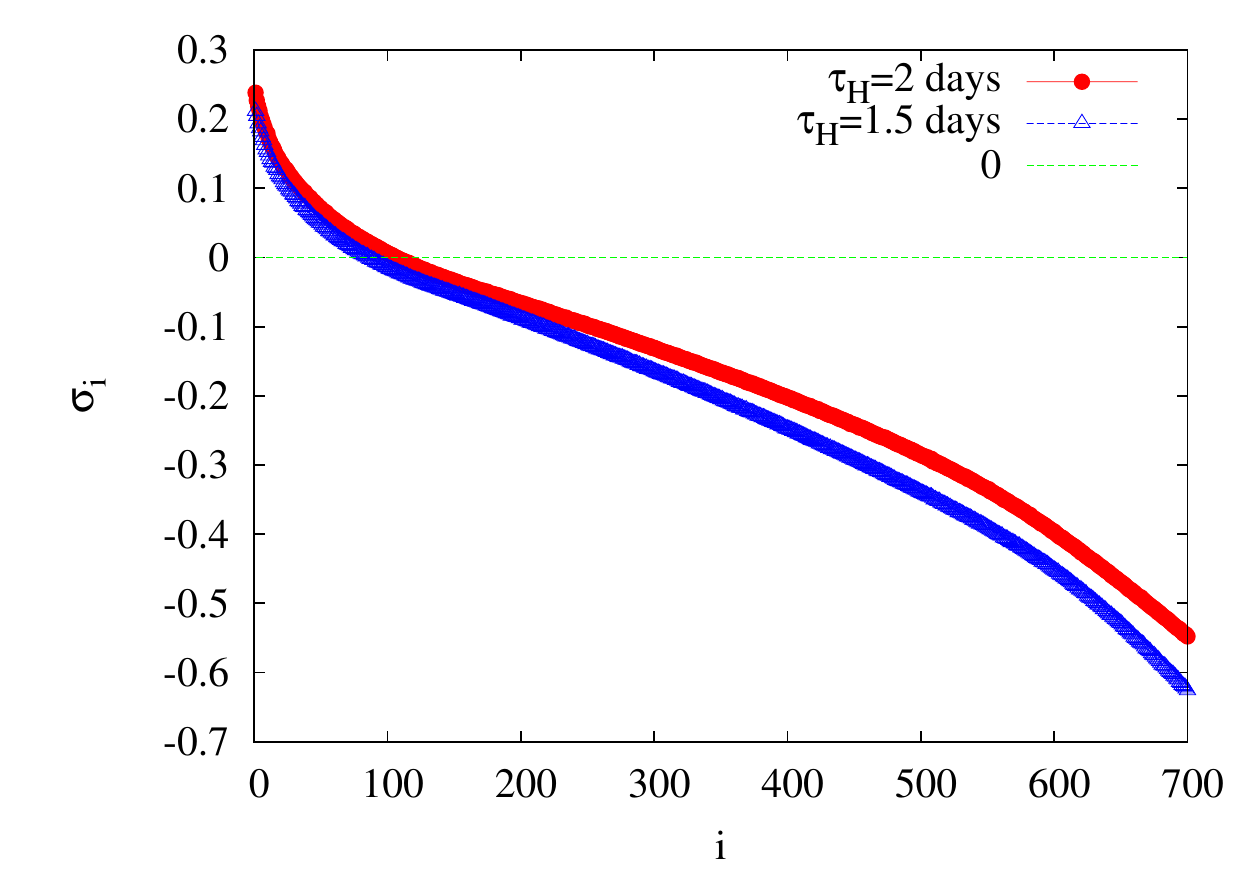}

\caption{
\label{QGspectra}
Lyapunov spectra (the 700 first exponents) of the atmospheric model described in Section \ref{QG3}, 
as obtained after an integration of 3,000 days in perpetual winter conditions.
Two different dissipation timescales were used, $\tau_H=2$ days (red filled circles) and $\tau_H=1.5$ days (blue triangles).}
\end{figure*}

In Fig. \ref{QGspectra} the first 700 Lyapunov exponents
obtained after 3,000 days of integration of the standard version of the model (\ref{vorticity}) with $\tau_H=2$ days, are displayed.
For a value of the dissipation timescale of $\tau_H=2$ days controlling the scale selective dissipation in the model 
(see Appendix A of \cite{Marshall1993}), the first 102 exponents are positive, 
the 103th is very close to zero and the next ones are negative (red filled circles in Fig. \ref{QGspectra}).
This result shows that the QG model lives on a high-dimensional chaotic attractor
displaying sensitivity to initial conditions. Furthermore, in view of the large number of close
positive exponents, it suggests that the Lyapunov spectrum is practically continuous.
The amplitude of the first exponent is equal to 0.23 days$^{-1}$ corresponding to a doubling time
of small errors of the order of $\ln (2/0.23) \approx 3$ days, a realistic order of magnitude for the error doubling time
in more sophisticated models at large scales \cite[e.g.][]{Yoden2007}. 

The second curve (blue triangles) in Fig. \ref{QGspectra} displays the Lyapunov spectrum for a smaller value of $\tau_H=1.5$ days,
inducing a higher dissipation in the atmospheric model, see \cite{Marshall1993, Vannitsem1997}. In this case the
number of positive exponents is reduced but the overall structure of the spectrum remains the same. 

\begin{figure*}
\includegraphics{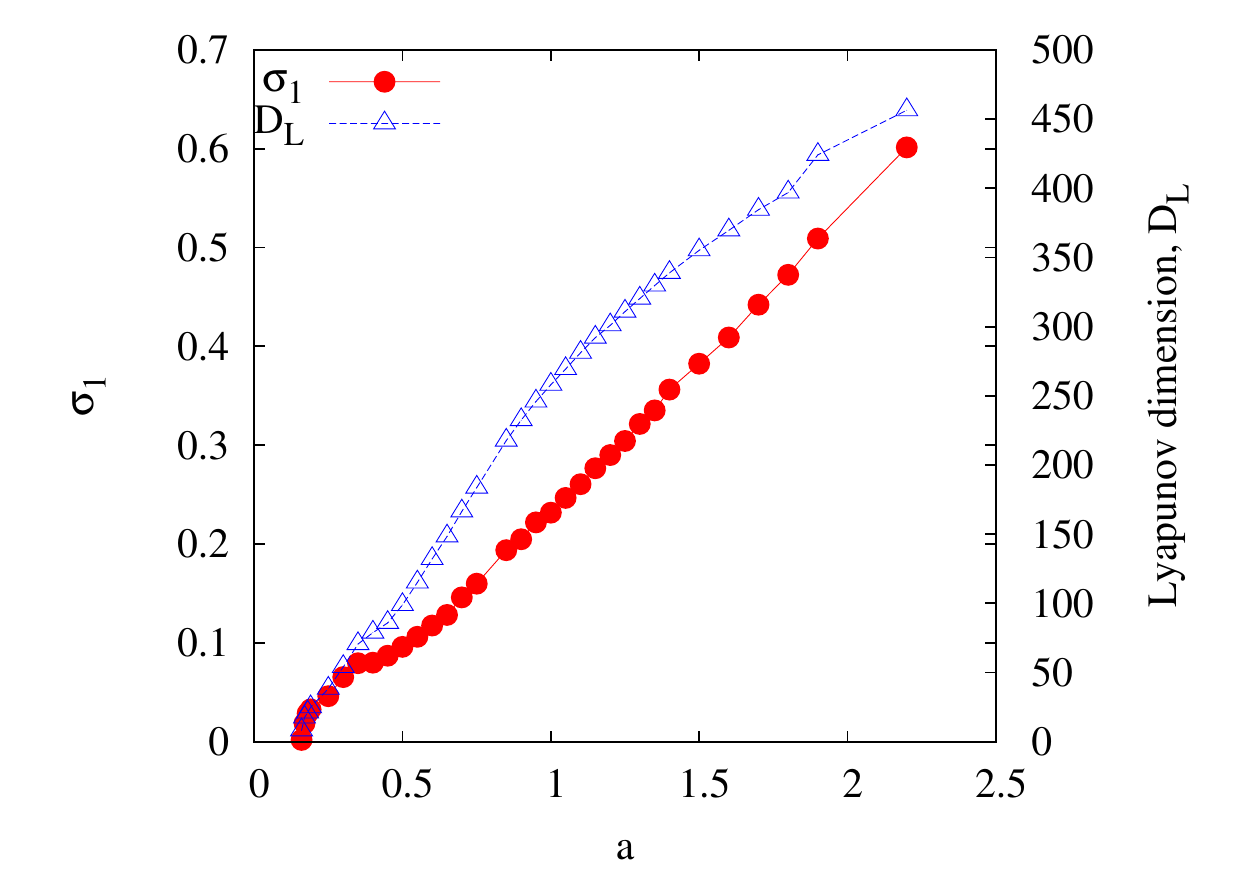}
\caption{
\label{QG1st}
Dependences of the dominant Lyapunov exponent, $\alpha_1$ (red filled circles), and of the Lyapunov dimension, $D_L$ (blue triangles), 
as a function of the coefficient $a$, multiplying the forcing, $a S_i$, of the atmospheric model (\ref{vorticity}). }
\end{figure*}

In order to figure out what is the sensitivity of the Lyapunov properties as a function of the
forcing, $S_i$, a multiplicative coefficient, $a$, is introduced in the model equation (\ref{vorticity}) as $a S_i$.   
Figure \ref{QG1st} displays the variations of the amplitude of the dominant exponent and of the
Lyapunov dimension as a function of $a$, as obtained with a set of experiments of 6,000 days of
integrations. {\sv The Lyapunov dimension is defined as,
$$D_{\rm{L}} = j^* + \frac{\sum_{j = 1}^{j^*} \sigma_j}{| \sigma_{j^*+1} |}, $$
where $j^*$ is the largest $j$ such that  $\sum_j \sigma_j > 0$.} 

The dependence of the Lyapunov instability properties as a function of $a$ is smooth with an
increase of the Lyapunov dimension up to about 450. This smoothness contrasts with the one found in low-order
models, and in particular with the CS model discussed in Section \ref{lyap_CSmodel}, but is in agreement
with the results highlighted recently in \cite{Lucarini2007, Schubert2015} in other intermediate order models.    

\begin{figure*}
\includegraphics{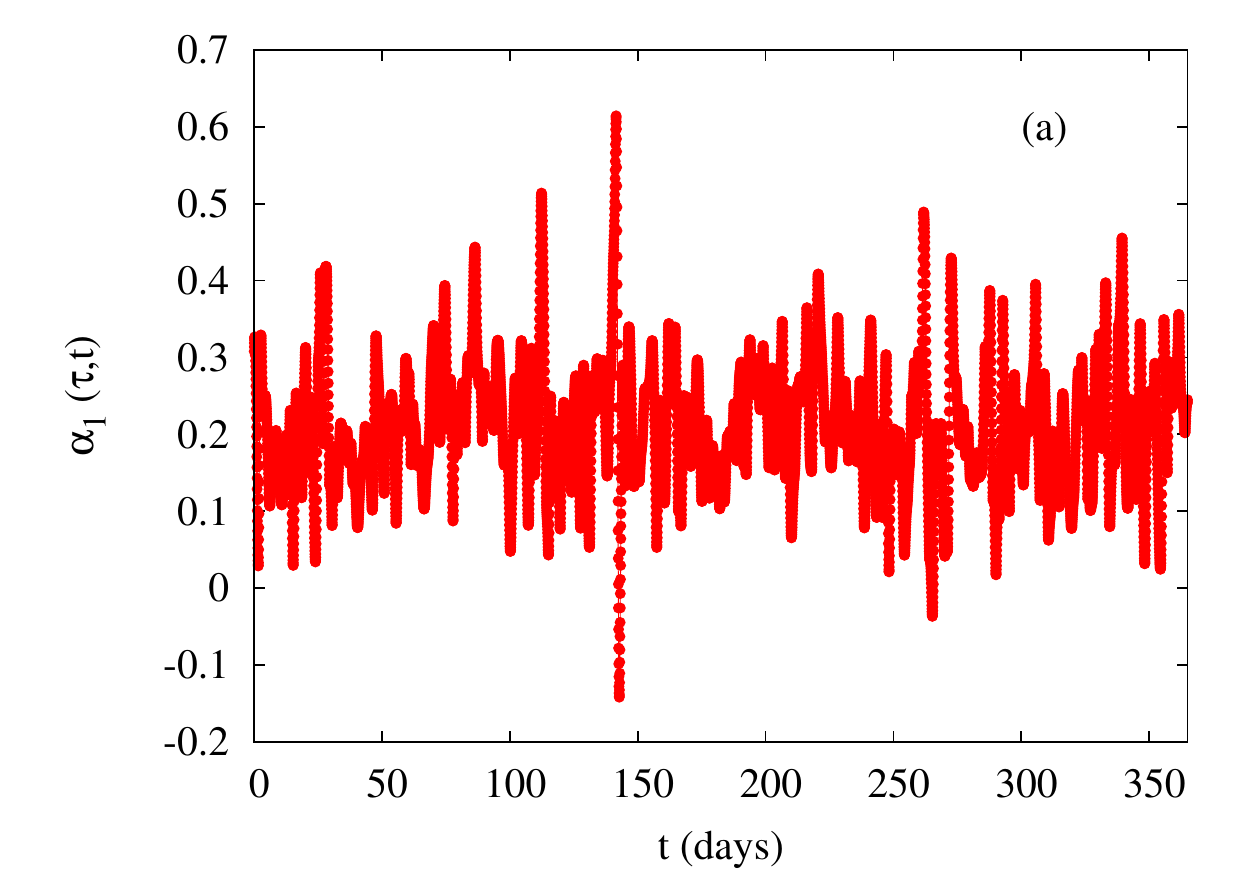}
\includegraphics{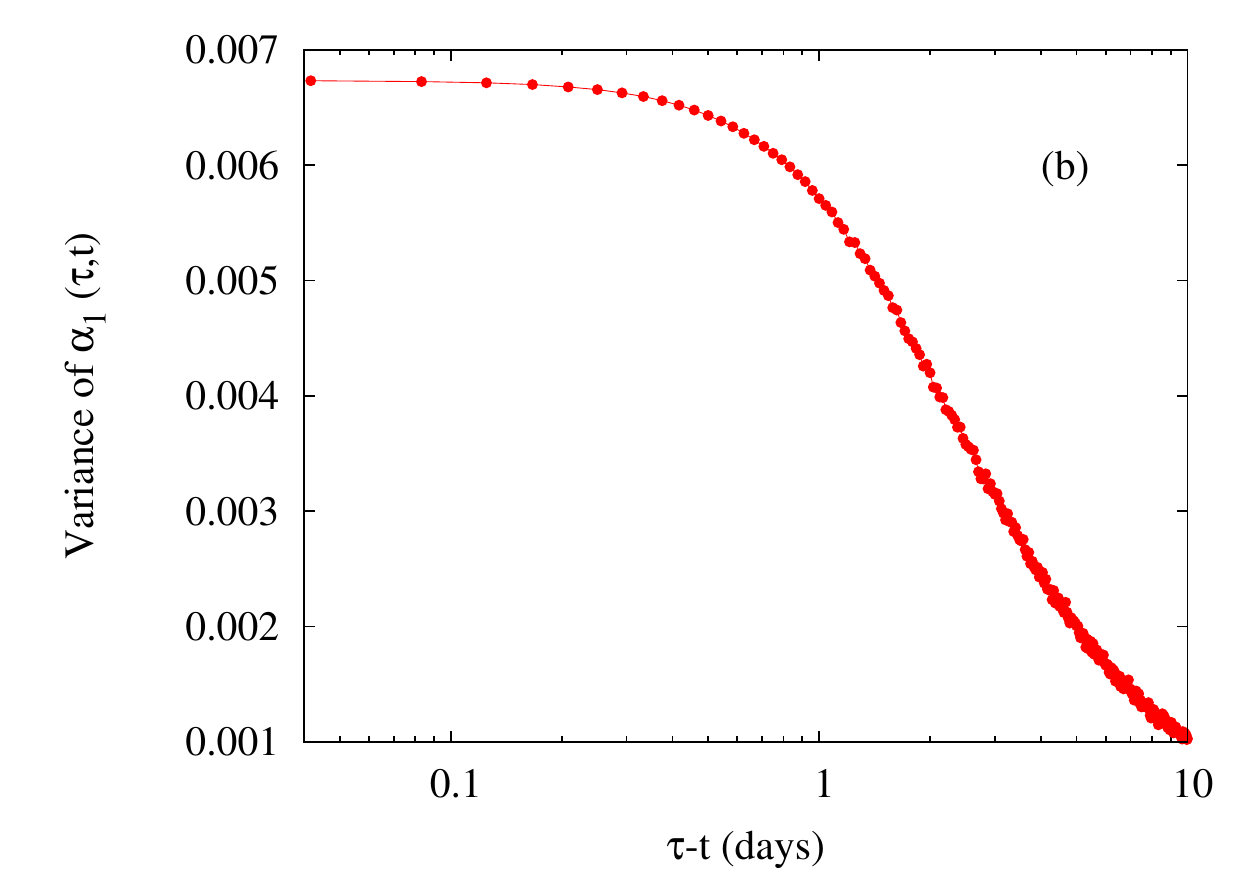}
\caption{
\label{QG1stloc}
(a) Temporal evolution of local amplification rate $\alpha_1(\tau,t)$ for $\tau-t=1$ hour; (b) Variance of 
$\alpha_1(\tau,t)$ as a function of $\tau-t$.}
\end{figure*}

The variability of the first exponent is represented in Fig. \ref{QG1stloc}a. The variability
of the dominant exponent is now mainly confined to positive values except in rare occasions. This 
variability contrasts with the one found in the CS model for which the variability is
much larger. In order to check whether this variability is due to the specific sampling chosen, we
have computed the variability of $\alpha_1 (\tau,t)$ as a function of $\tau-t$ (Fig. \ref{QG1stloc}b), as suggested in 
\cite{Trevisan1998}. 
This variability seems already close to convergence when $\tau-t$ is of the order of 1 hour. One
can therefore conclude that the variability of the dominant Lyapunov exponent is  weaker in the 
intermediate order atmospheric model than in the low-order CS model. 

These results highlight the contrast of the instability properties between low-order and intermediate order
atmospheric models. In particular, a decreased {\sv variability} of the local Lyapunov instabilities on the attractor
of the model is observed. This feature is opening an important question to know whether this variability 
still decreases when the number of variables is 
further increased. The question is  closely related to the open problem of the effective hyperbolicity 
(or partial hyperbolicity) of high-dimensional systems as discussed in \cite{Gallavotti1995, Vannitsem2016}.   

This question is not purely academic but could have important implications for operational
weather forecasts since a variability of the local instability properties of high resolution
models is experienced \cite{Ehrendorfer1997, Yoden2007}. Is this variability already present at the level of the large scale 
dynamics of 
atmospheric flows (as described by the quasi-geostrophic equations) or rather to processes that are not represented in this type of model,
such as large scale divergent flows, convection, precipitation, gravity waves,... interfering with the large scale     
dynamics?

\subsection{Lyapunov exponents of the low-order coupled ocean-atmosphere system \label{OA-lyap}}

The atmosphere is also subject to boundary forcings coming from the other components
of the climate systems that could presumably affect its predictability properties. Obvious candidates are the oceans 
that are interacting with the atmosphere through exchanges of momentum, mass, heat and radiations. This  question is now
addressed through the analysis of a low-order coupled ocean-atmosphere model described in Section \ref{low-order}, see also 
\cite{Vannitsem2015a, Vannitsem2015b}.

\begin{figure*}
\includegraphics{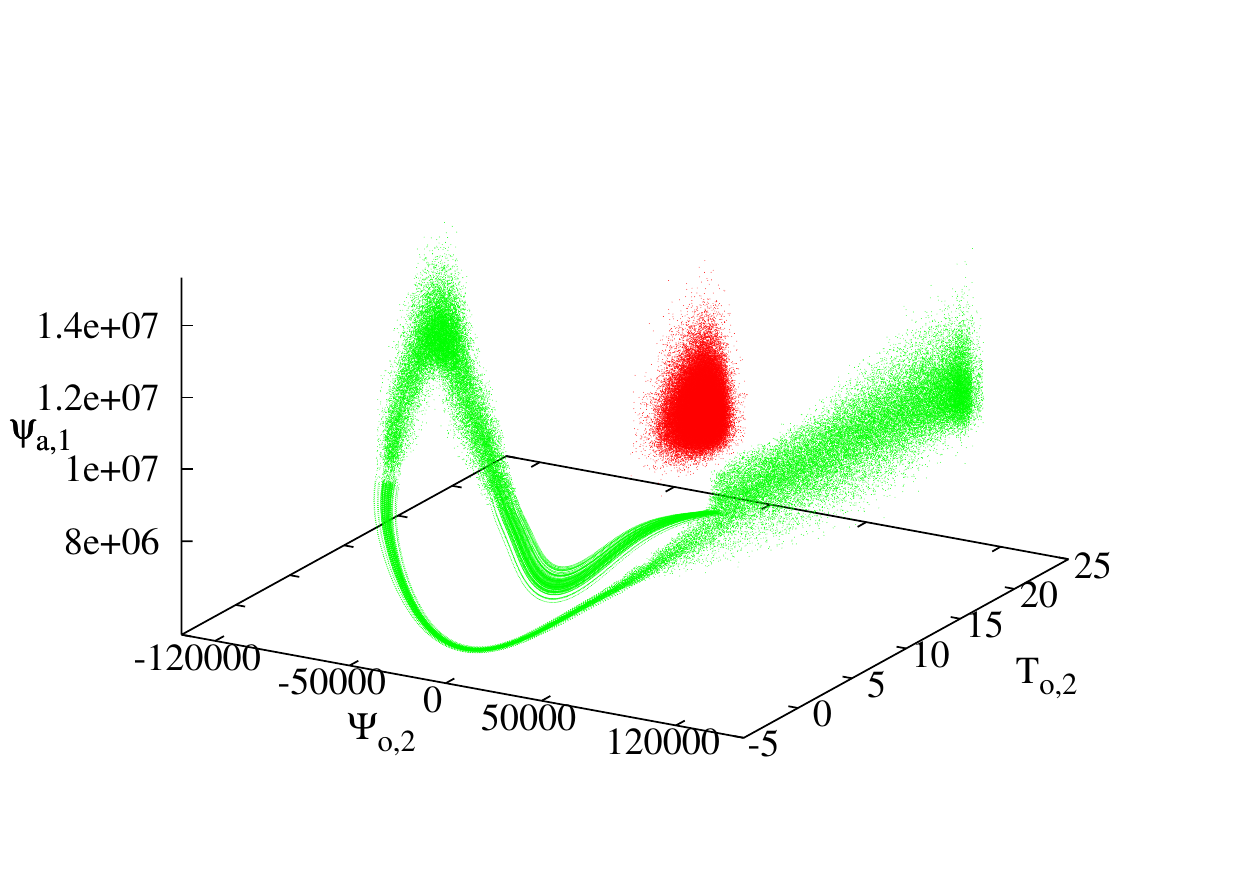}
\caption{
\label{OAattractor}
Three-dimensional projection $(\psi_{{\rm a},1}, \psi_{{\rm o},2}, T_{{\rm o},2})$ of the attractors of the solutions of the low-order coupled ocean-atmosphere model described in 
section \ref{low-order}, for parameters $S_o=310$ W m$^{-2}$, $D_{ref}=100$ m, with two different values of $C=0.010$ 
(red dots) and $C=0.015$ kg m$^{-2}$ s$^{-1}$ (green dots). These attractors will be referred to as the red and green attractors, respectively. }
\end{figure*}

The three-dimensional projections along the variables $(\psi_{{\rm a},1}, \psi_{{\rm o},2}, T_{{\rm o},2})$ 
of the attractors are illustrated in Fig. \ref{OAattractor}
for two different values of $C=0.010$ and $C=0.015$ kg m$^{-2}$ s$^{-1}$. 
These attractors show fundamentally different properties, one of them displaying 
a dynamics around a well defined unstable periodic orbit identified in \cite{Vannitsem2015b} (green dots). 
As discussed in details in \cite{Vannitsem2015b}
the development of the attractor around this unstable orbit is inducing a low-frequency variability on decadal 
timescales and allows for long term predictions beyond the usual 10-15 days weather forecasts. This point will be further
discussed in Section \ref{error}. 

\begin{figure*}
\includegraphics{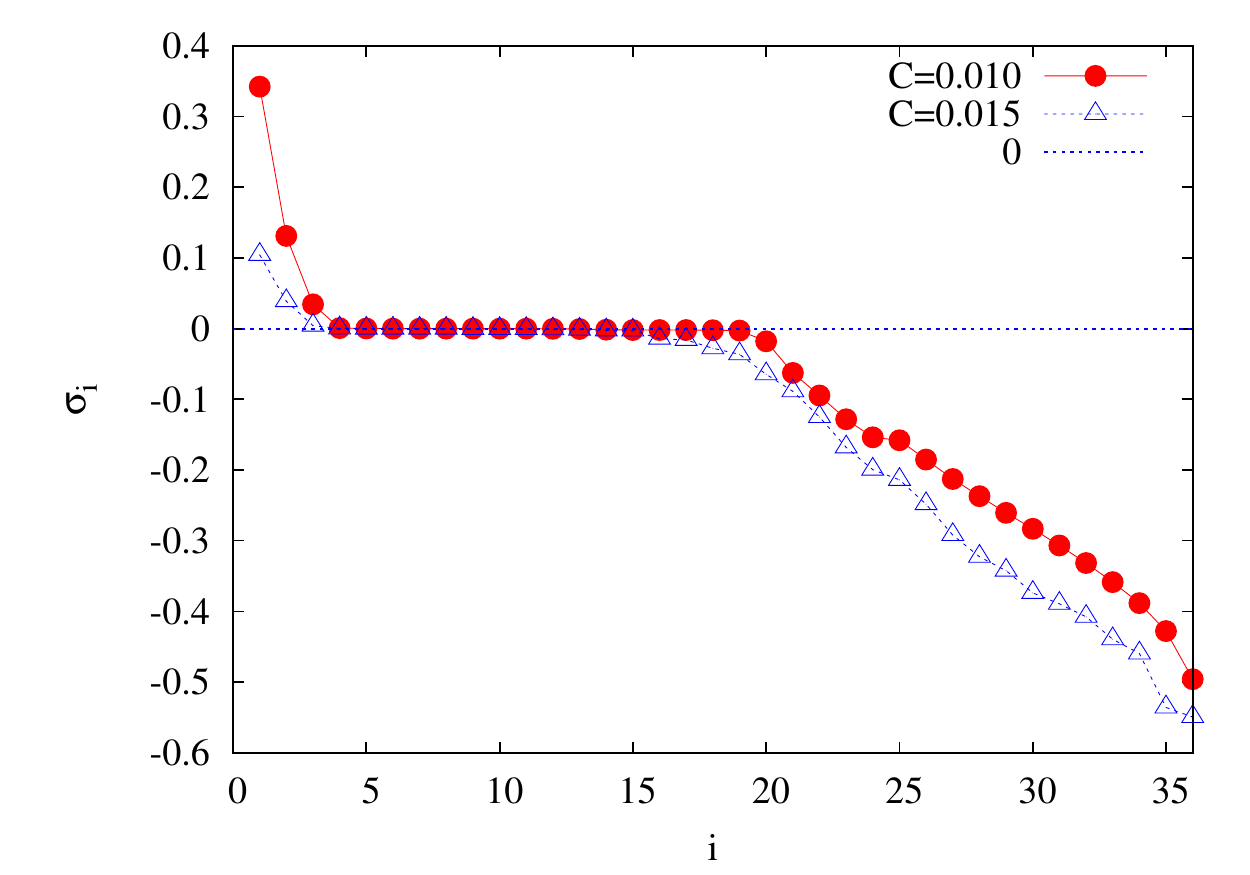}
\caption{
\label{OAspectra}
Lyapunov spectra for the red (red filled circles) and green (blue triangles) attractors displayed in Fig. \ref{OAattractor}.
}
\end{figure*}

Figure \ref{OAspectra} displays the Lyapunov spectra for the two attractors. A first remarkable result is
the presence of a large set of Lyapunov exponents close to 0. These are associated with the presence of the ocean
whose typical dissipative timescale, $1/r$, is much longer than for  
the atmosphere, as discussed in \cite{Vannitsem2016}. {\sv The Covariant Lyapunov vectors associated with this group
of exponents display angles with the tangent vector to the trajectory that are small (as compared to the other vectors)}, and 
form a near-neutral manifold of high dimension in which the 
error amplification (or contraction) is small. The presence of a large number of near zero exponents also implies that the 
Lyapunov dimension, $D_L$, of the system is large, even if the dimension of the unstable subspace is small. 
This feature may in particular have an important impact on the development of data assimilation
schemes exploiting the separation of stable and unstable-neutral manifolds as proposed in \cite{Trevisan2004, Carrassi2007, Trevisan2011}.  

Another important result is the
small amplitude of the positive exponents when the low-frequency variability is developing in the system 
(green attractor of Fig. \ref{OAattractor}). In this case the system is stabler due to the strong influence of the ocean dynamics. 
In the climate community these two types of dynamics are usually {\sv referred to} as
{\it passive} or {\it active} ocean dynamics \cite{Vanveen2003}. In the dynamical systems framework these 
qualitative changes of dynamics is explained through a bifurcation from which new types of solutions
are emerging, that would not be present without the ocean-atmosphere coupling \cite{Vannitsem2015b}.
This qualitative change of dynamics has also considerable implications for the predictability of the coupled 
ocean-atmosphere system. This important aspect will again be addressed in Section \ref{error} where the
error dynamics is discussed.

\begin{figure*}
\includegraphics[width=0.45\textwidth]{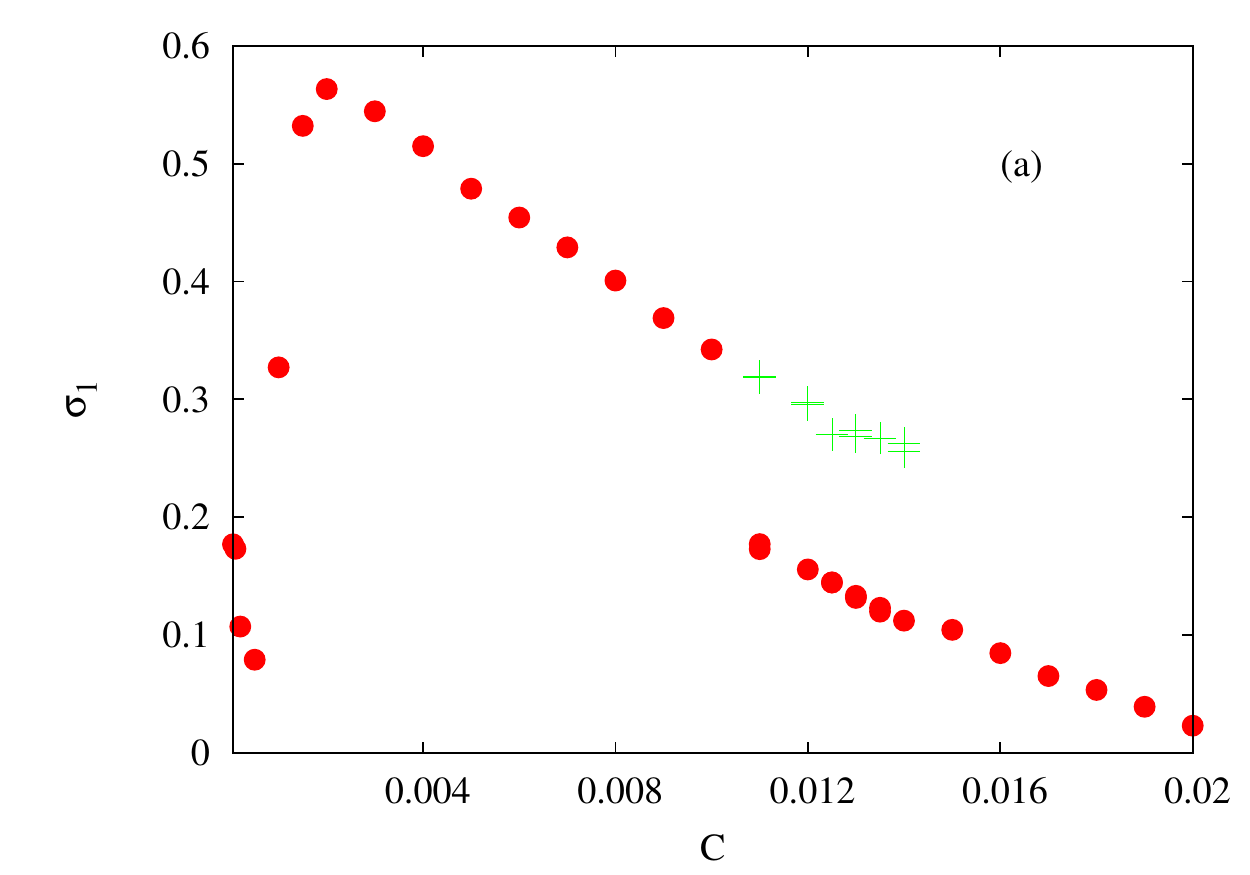}
\includegraphics[width=0.45\textwidth]{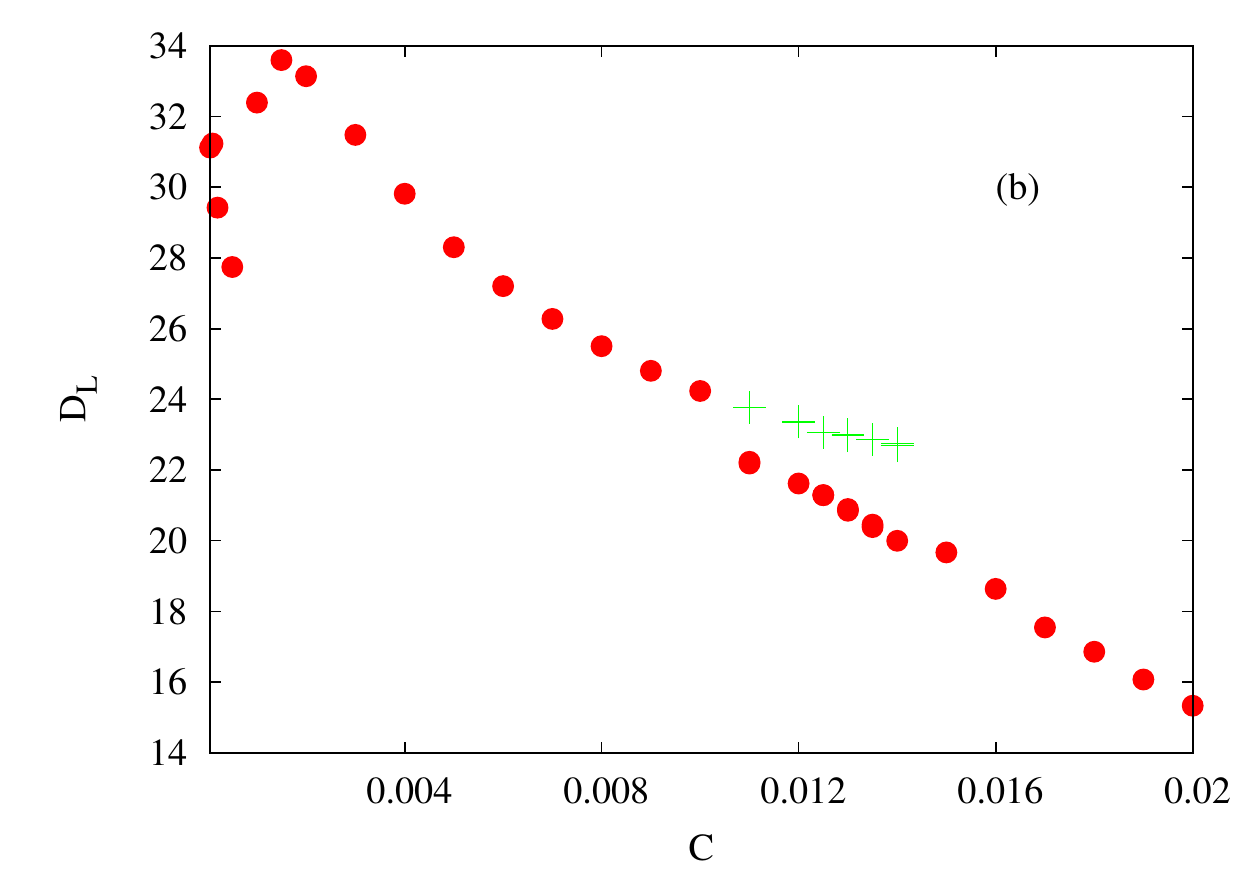}

\includegraphics[width=0.45\textwidth]{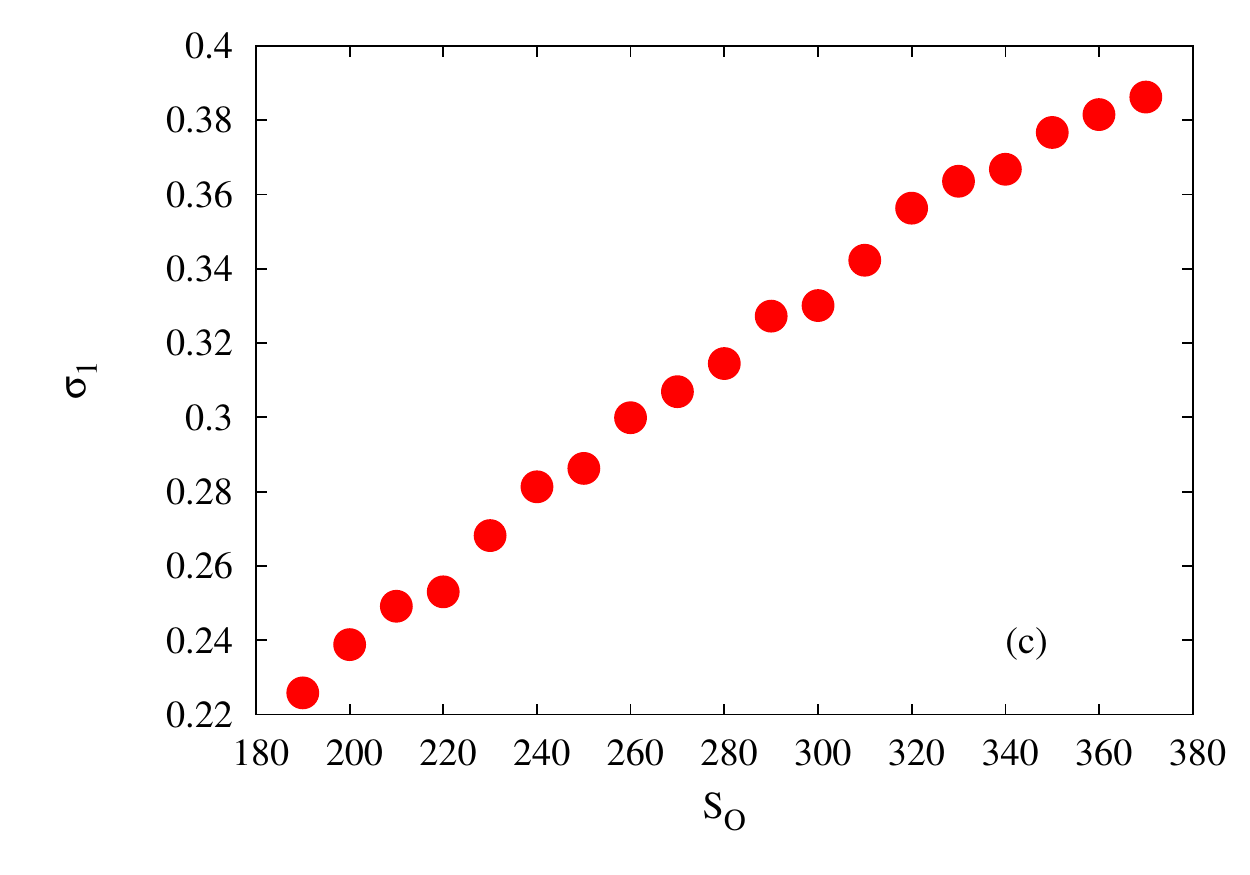}
\includegraphics[width=0.45\textwidth]{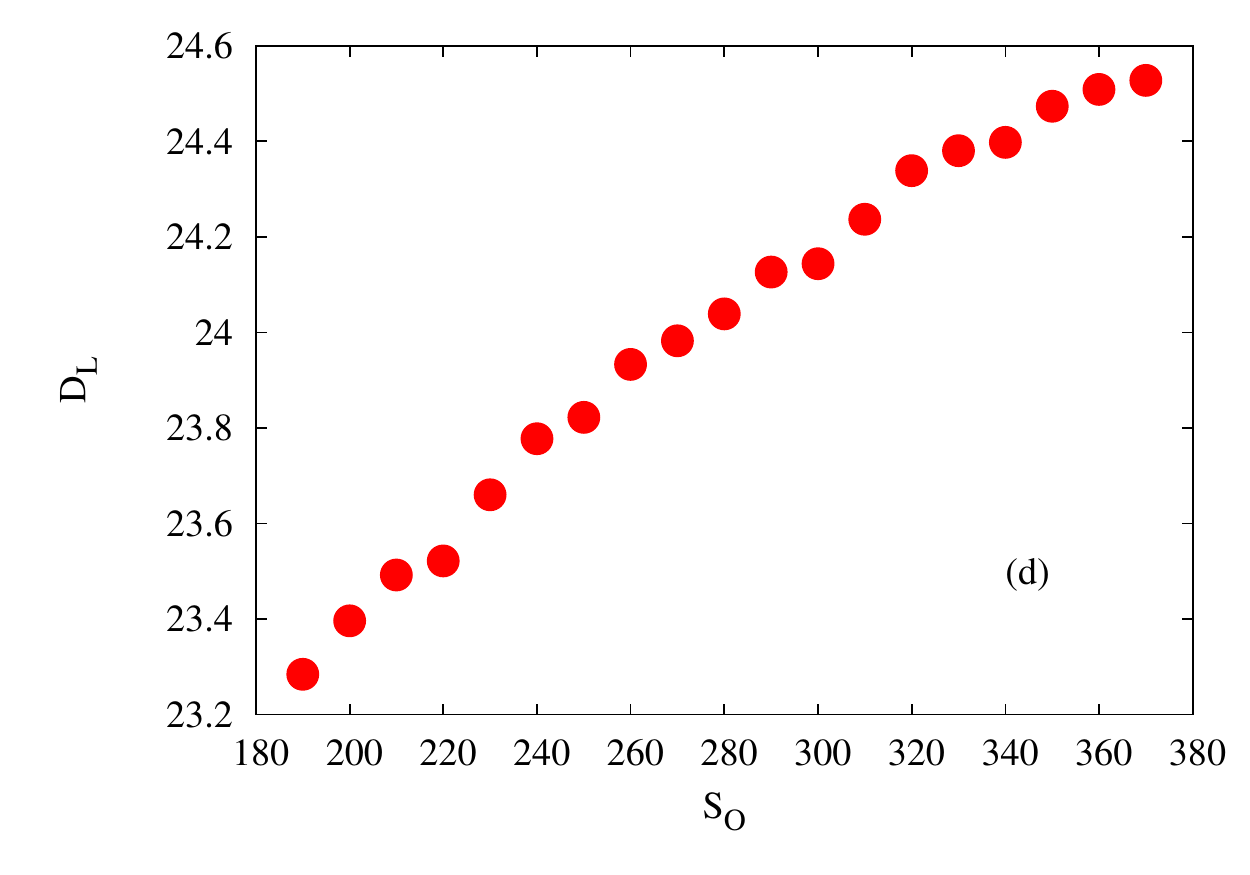}
\caption{
\label{OA1st}
First Lyapunov exponent (a) and Lyapunov dimension (b) as a function of the friction parameter $C$ for the coupled ocean-atmosphere model with
 $S_o=310$ W m$^{-2}$, $D_{ref}$=100 m and $\kappa=1$. The green pluses represent the results obtained with long transient trajectories that have not
converged toward the final attractor yet, and living close to the red attractor of Fig. \ref{OAattractor}. (c) and (d) as (a) and (b) but as a function
of $S_o$ for $C=0.01$ kg m$^{-2}$ s$^{-1}$, $D_{ref}$=100 m and $\kappa=1$.  
}
\end{figure*}

Figures \ref{OA1st}a--b display the dependence of the 1st Lyapunov exponent and the Lyapunov dimension 
of the coupled system as a function of the surface friction coefficient, $C$, for one specific value of 
$S_o=310$ W m$^{-2}$.  For values of $C$ smaller than $0.011$ kg m$^{-2}$ s$^{-1}$, the solutions are converging toward
an apparently unique attractor, e.g. the red attractor of Fig. \ref{OAattractor}. 
 When $C$ is further increased and for a quite substantial range of $C=[0.011, 0.014]$ kg m$^{-2}$ s$^{-1}$, the 
system displays solutions that can show very long transients around the attracting set
present for smaller values of $C$, that eventually end up in a very different region of the phase space for which
the attracting set displays a shape similar to the green attractor of Fig. \ref{OAattractor}. The Lyapunov
spectra and dimensions were also computed for these long transient dynamics and displayed as green pluses in Figures \ref{OA1st}a--b.  
Beyond that range the convergence is faster and the attracting set of the attractor of the solution ressembles the green one displayed 
in Fig. \ref{OAattractor}.

A second result of interest is the presence of a maximum in the
amplitude of the dominant exponent and of the Lyapunov dimension when the surface friction coefficient 
is decreased.  Below a friction coefficient of about $0.0015$ kg m$^{-2}$ s$^{-1}$ the {\sv amplitude of the dominant Lyapunov exponent decreases}. 

In order to understand this result, one must first realize that when $C$ is decreased both the momentum and heat transfers
between the two sub-systems are decreased \cite{Vannitsem2015a}, reducing the coupling between the two systems and the dissipation
within the atmosphere. This implies that the instability properties of the flow dominated by the atmospheric dynamics is increased. 
When $C$ is becoming too small a qualitative change of dynamics seems to occur but the slow convergence of the trajectories does
not allow us to clarify the specific nature of the dynamics yet. This problem is left for a future study.   
 
Interestingly the variations of the dominant exponent and of the Lyapunov dimension in the range
of interest for climate modelling, $C=[0.005, 0.020]$ kg m$^{-2}$ s$^{-1}$, look smooth, except in the transition zone $[0.10,0.11]$. 
A similar picture can be drawn when
changing $S_o$ for a fixed value of $C$, as illustrated in Figs. \ref{OA1st}c--d, with smooth
variations of the Lyapunov instability properties as a function of the radiative forcing.
This result contrasts with the usual picture that can be drawn from very low-order
systems (typically of 3-4 variables) for which much more complicate bifurcation diagrams are obtained.
As discussed in \cite{Lucarini2007}, it seems to be a natural property when the phase space dimension
of the system increases. 

\begin{figure*}
\includegraphics{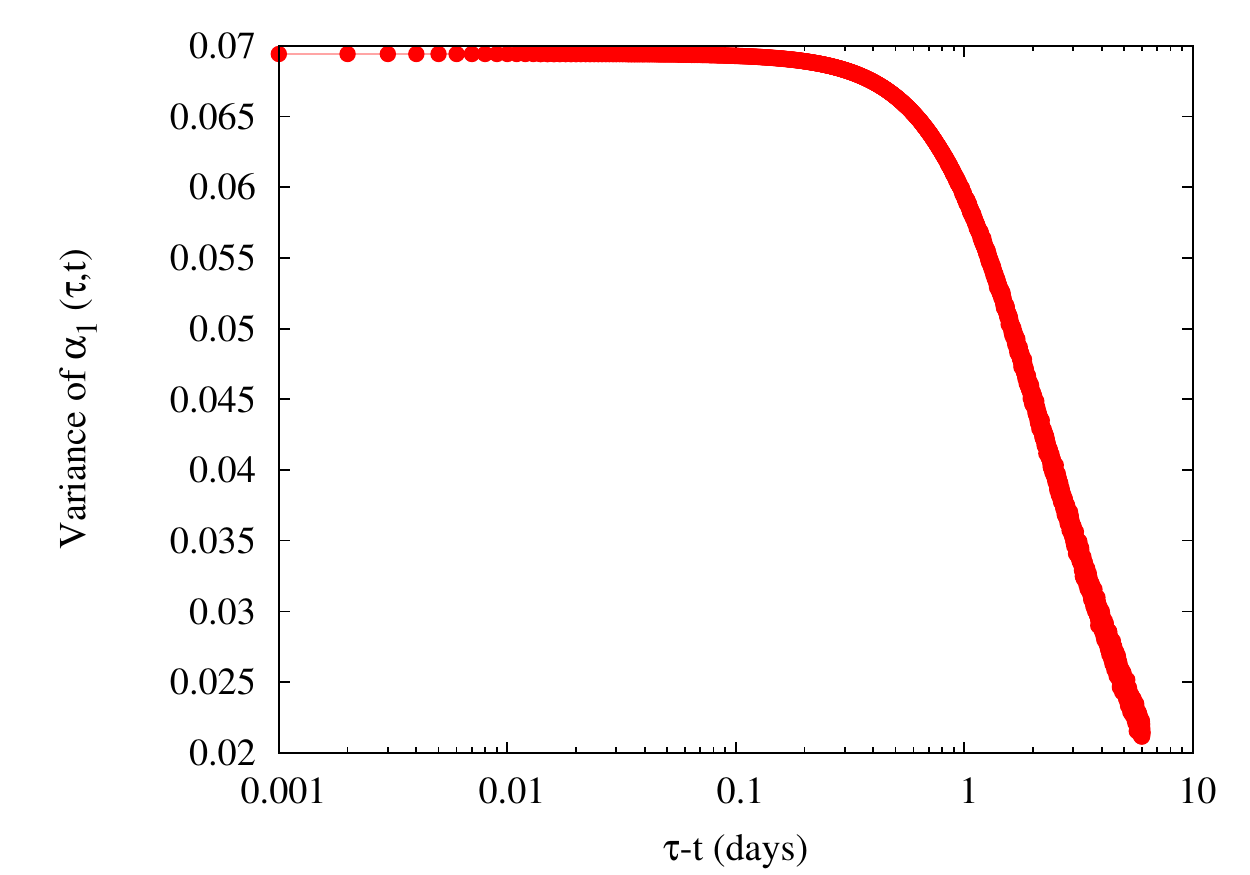}
\caption{
\label{OA1stloc}
Variance of $\alpha_1(\tau,t)$ as a function of $\tau-t$ for the ocean-atmosphere model, with parameters $S_o=310$ W m$^{-2}$, $D_{ref}$=100 m, 
$C=0.01$ kg m$^{-2}$ s$^{-1}$ and $\kappa=1$.
}
\end{figure*}

Figure \ref{OA1stloc} illustrates the variance of $\alpha_1(\tau,t)$ as a function of $\tau-t$ for the ocean-atmosphere
coupled model. A picture intermediate between the results obtained with the CS model and the QG3T21 model emerges,
with a variance of $\alpha_1(\tau,t)$ converging toward a value of about $0.07$ day$^{-2}$ for $\tau-t \rightarrow 0$.

Up to now, the focus was put on the autonomous version of the low-order models but the
atmosphere is strongly influenced by the natural seasonal variability. This can be taken into account 
in the present model by introducing realistic seasonal variations as discussed in Section \ref{low-order}.  

\begin{figure*}
\includegraphics{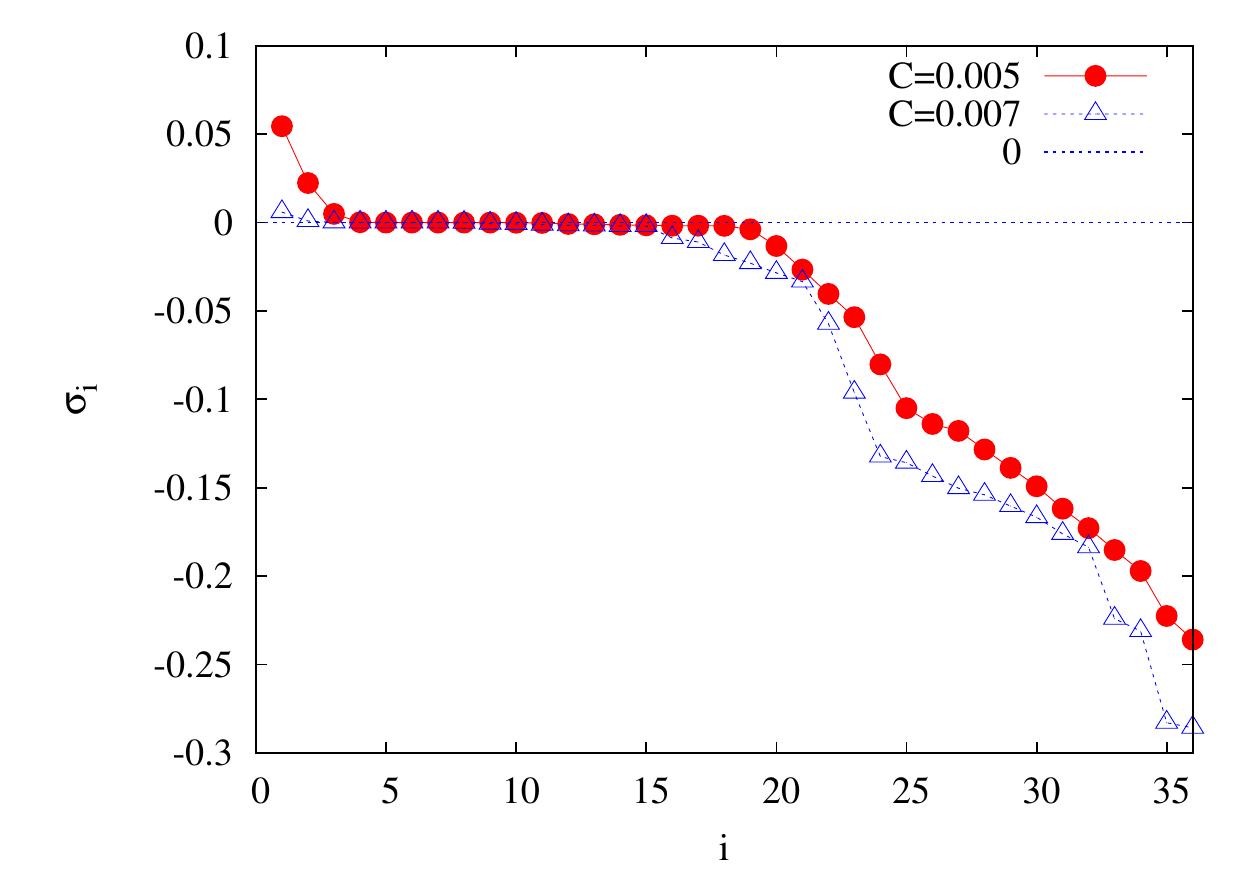}
\caption{
\label{OAspectra-naut}
Lyapunov spectra as obtained with two different values of the friction parameter $C=0.007$ (blue trianges) and $C=0.005$ (red filled circles) 
kg m$^{-2}$ s$^{-1}$, corresponding to attractors with and without low-frequency variability, respectively. 
The other parameters used are $S_o=310$ W m$^{-2}$, $D_{ref}$=100 m and $\kappa=0.3$.
}
\end{figure*}

Figure \ref{OAspectra-naut} shows the Lyapunov spectra as obtained with two different values
of the friction parameter $C=0.007$ and $C=0.005$ kg m$^{-2}$ s$^{-1}$, corresponding to attractors with and without low-frequency
variability respectively. Both attactors are chaotic but the amplitude of the dominant Lyapunov
exponent is relatively small in both cases. The main reason is the fact that when the seasonality
is imposed the meridional gradient of radiative input has a smaller amplitude than in the autonomous case
considered above whatever is the time of the year. 

\begin{figure*}
\includegraphics[width=0.45\textwidth]{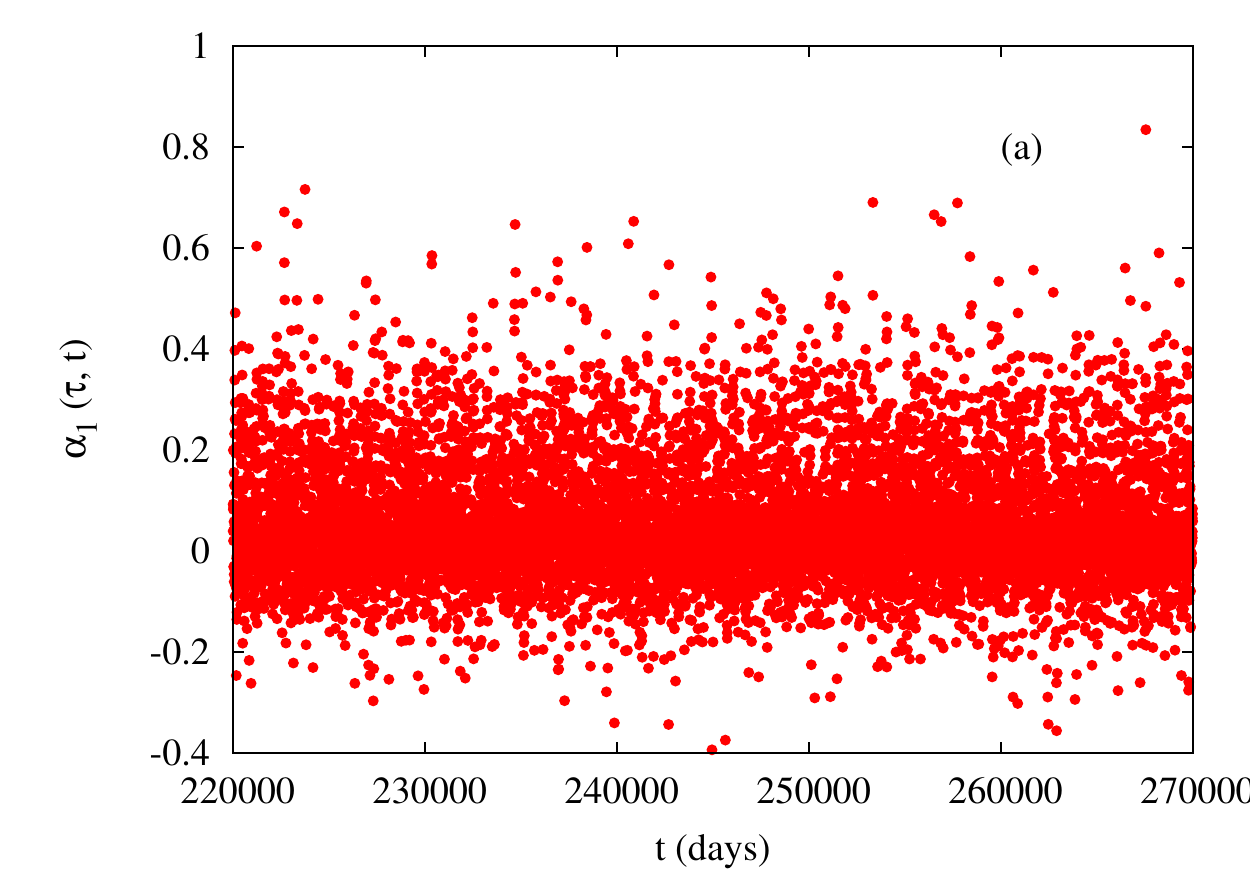}
\includegraphics[width=0.45\textwidth]{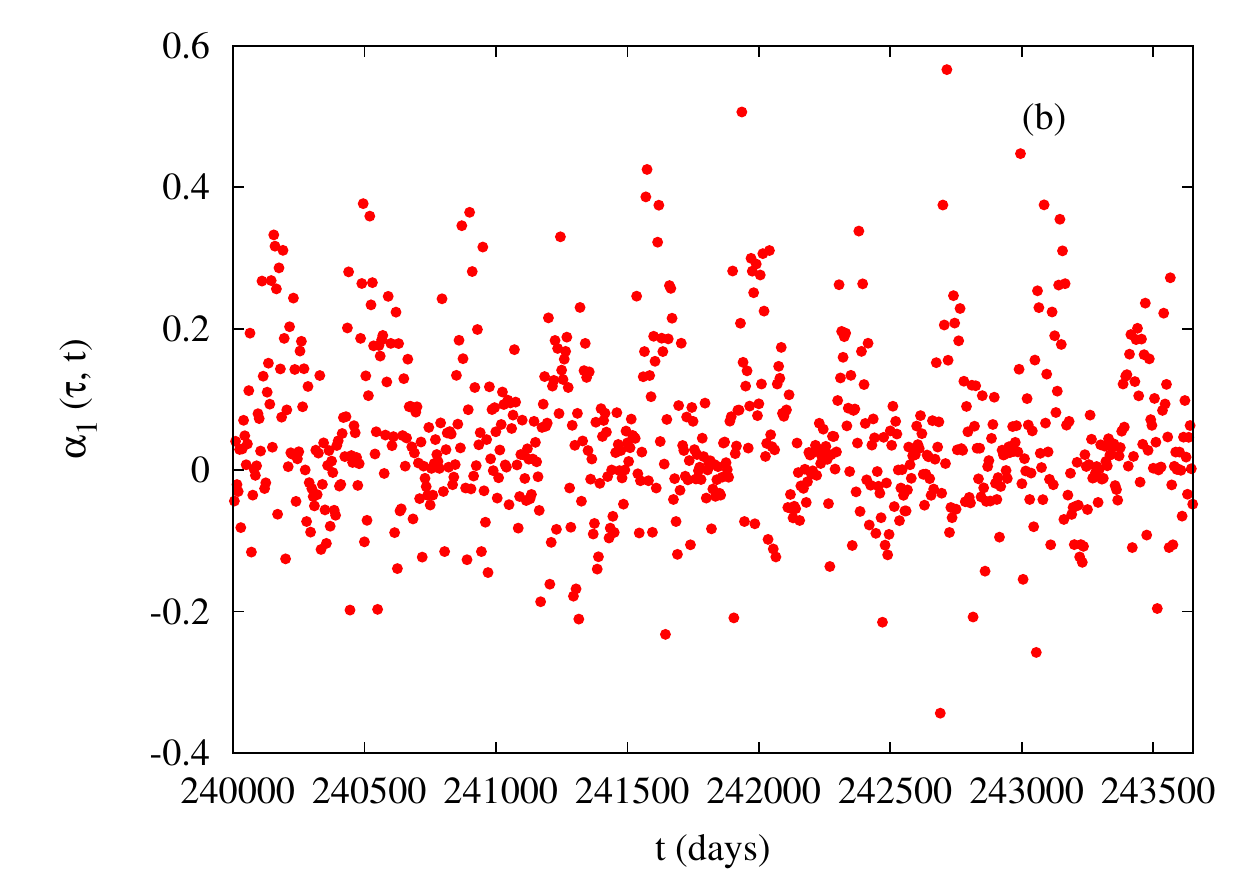}

\includegraphics[width=0.45\textwidth]{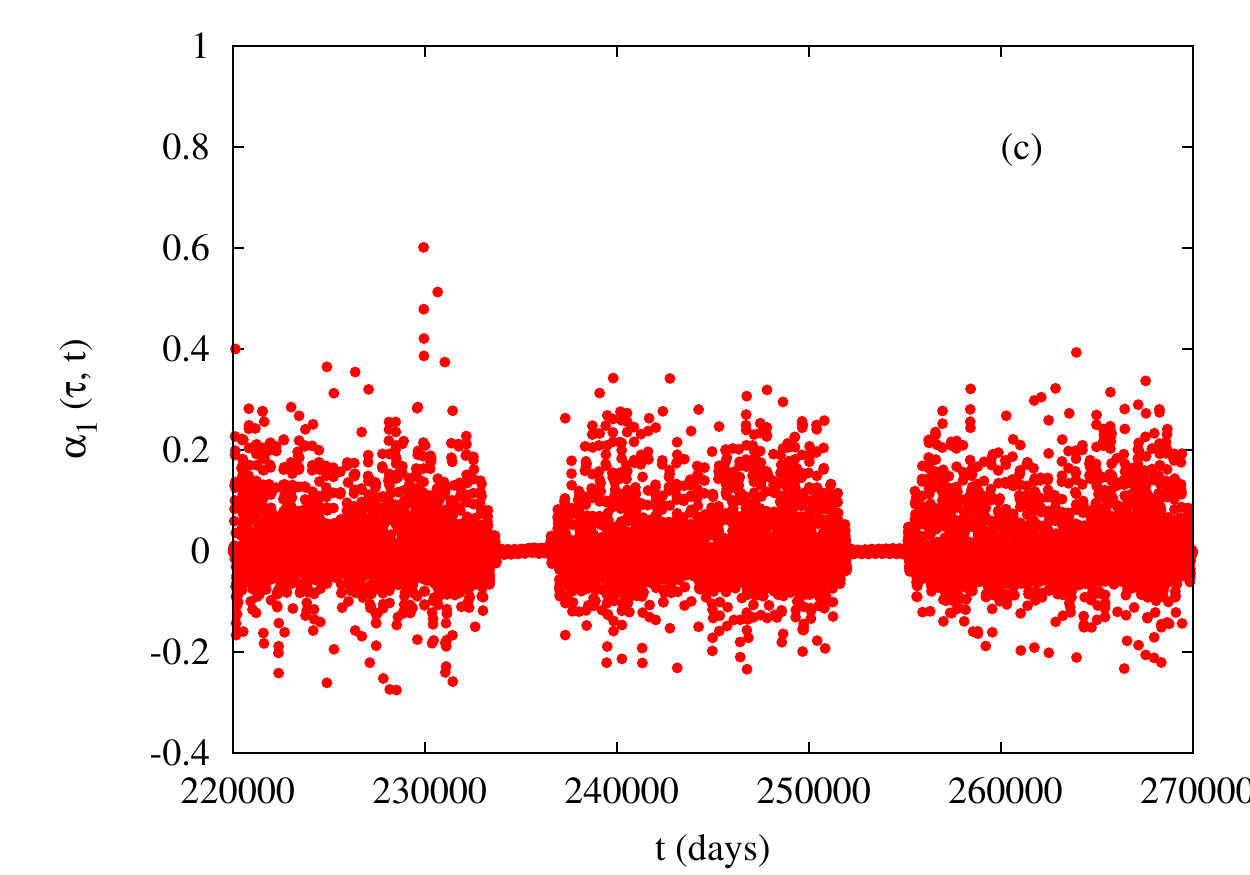}
\includegraphics[width=0.45\textwidth]{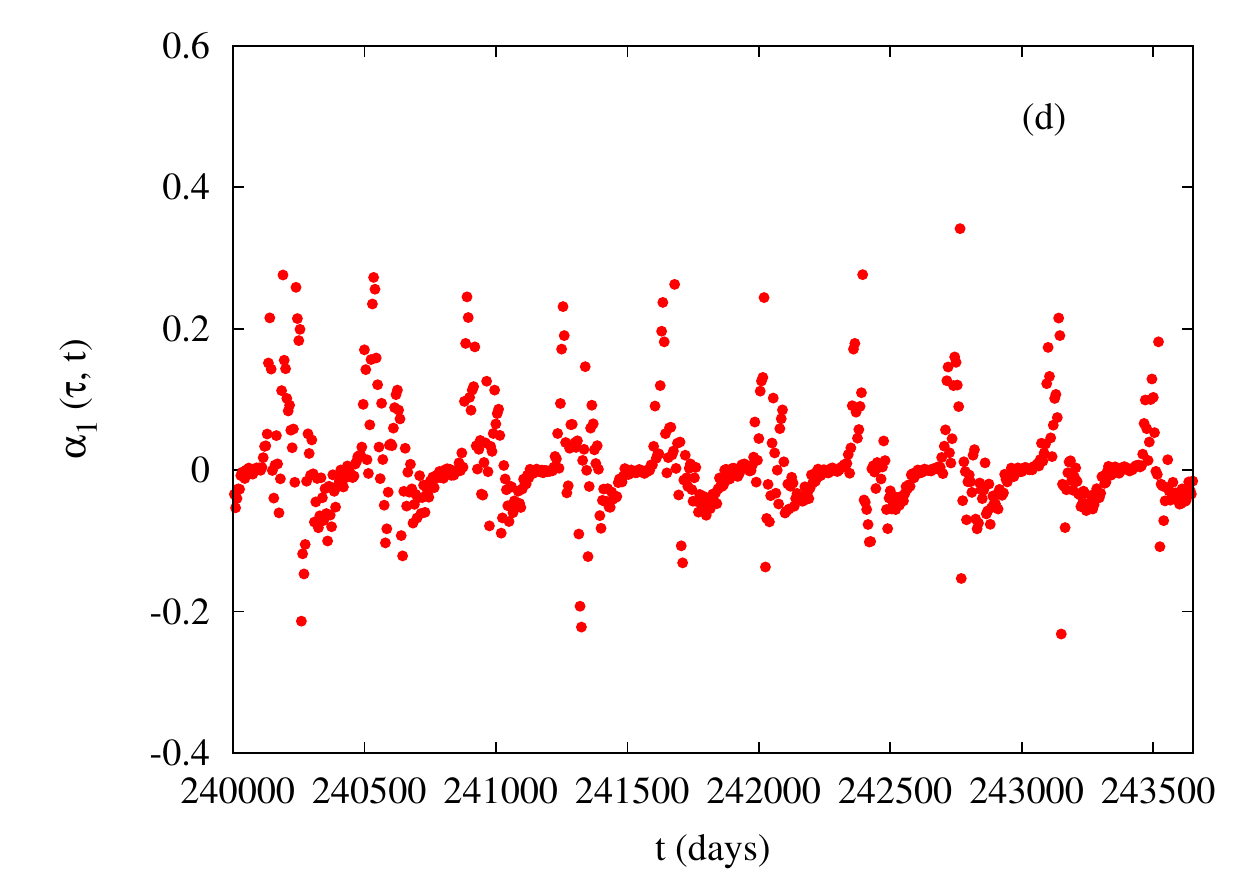}
\caption{
\label{OA1st-naut}
Temporal variability of the local Lyapunov exponents, $\alpha_1(\tau,t)$, as obtained with
$\tau-t=0.005$ days and sampled every 5 days, for the two
parameter values (a) $C=0.005$ and (c) $C=0.007$ kg m$^{-2}$ s$^{-1}$. A zoom on a 10-year time series is also provided  
at panels (b) and (d) for the two parameter values. The other parameters used are $S_o=310$ W m$^{-2}$, $D_{ref}$=100 m and $\kappa=0.3$.
}
\end{figure*}

A more interesting finding is the temporal variability of the local Lyapunov exponents, $\alpha_1(\tau,t)$ with 
$\tau-t=0.005$ days, sampled every 5 days. These are displayed in Fig. \ref{OA1st-naut} for the two
parameter values $C=0.005$ (a) and $C=0.007$ (c)  kg m$^{-2}$ s$^{-1}$. A zoom on a 10-year time series is also provided  
at panels (b) and (d) for the two parameter values. For $C=0.007$ kg m$^{-2}$ s$^{-1}$, the local Lyapunov exponents display large modifications
of their variability on a timescale of about 20,000 days with long quiescent periods as already
found in the autonomous version when the low-frequency variability is setting up, contrasting with the results obtained
with $C=0.005$. When zooming in on a 10-year
period (panels (b) and (d)) a seasonal signal is clearly visible with low local instabilities in summer
and high in winter. This result is in agreement with the common view that the weather (large-scale flow pattern) 
is more predictable in summer than in winter at mid-latitudes in the Northern hemisphere.  

{\sv In summary} the results presented above reveals a complicate picture of the instability properties of
multi-scale (autonomous and non-autonomous) systems. As the multi-scale nature of the dynamics is ubiquitous in the climate system 
(and in environmental modelling in general), we suspect that such properties are generic in the real world.

\section{Error dynamics \label{error}}

As mentioned in the Introduction, the growth of small initial errors arising from the finite
precision of the observational data and of the process of data-assimilation, is an intrinsic
property of atmospheric flows. It introduces irreducible limitations in the
possibility to forecast its future states beyond a predictability horizon, which may depend on the
type and scale of the phenomenon under consideration. 
We turn now on the analysis of the error dynamics in the hierarchy of models introduced in Section
II. 

\subsection{Error dynamics: generalities}

As mentioned in Section \ref{theory} the Lyapunov exponents characterizing the predictability of 
chaotic systems are defined in the limit of infinitely small errors and infinitely long times. In 
reality,  these limits are never reached and one must investigate the dynamics
of finite-size initial errors on a finite-time horizon. One starts with the definition already
introduced at Eq. (\ref{eucl}) in which the classical L2 norm is used. Since one is dealing
with finite-size errors, one must perform an ensemble average over the attractor of the system, 
\begin{eqnarray}
\langle E_t^2\rangle & = & \int d\vec{\epsilon}_0 \rho_{\epsilon} (\vec{\epsilon}_0) \nonumber \\
 \int d\vec{x}_0 & & \rho_x(\vec{x}_0) (\vec{x}'(t) - \vec{x}(t))^T {\bf{K}} (\vec{x}'(t) - \vec{x}(t)) \nonumber \\
\label{L2norm}
\end{eqnarray}
where $\rho_{\epsilon} (\vec{\epsilon}_0)$ and $\rho_x(\vec{x}_0)$, are the invariant probability distribution of the initial errors and 
of the initial conditions on the attractor of the system (provided that the system is ergodic). A matrix ${\bf{K}}$ is introduced in this relation allowing for choosing 
the specific norm of interest, for instance the energy norm or the enstrophy norm \cite[e.g.][]{Vannitsem1997}. Numerically the error evolution can be
evaluated by sampling a large set of initial conditions along a reference trajectory running along the ergodic attractor of the system under consideration, 
and to perform additional integrations starting from slightly perturbed initial conditions.   

In the following, the amplitude of the perturbations $\vec{\epsilon}$ is taken
sufficiently small in order to get information on the different regimes of error growth. In this case three main regimes are expected,
an exponential-like, a linear and  a saturation regime, see \cite{Nicolis1995, Vannitsem1994}. This dynamics can be empirically described by a simple logistic
law of the form \cite{Lorenz1982},
\begin{equation}
\frac{dE}{dt}=aE-bE^2
\label{logis}
\end{equation}
where E is the mean amplitude of the distance, e.g. (\ref{L2norm}), between two fields, and $a$ and $b$ some regression coefficients.
This description is, however, a rough approximation that does not take into account the natural variability of the local 
amplification rates along the attractor
of the system discussed in Section \ref{lyap}, nor the differential behavior of the error among spatial scales for which more 
detailed descriptions are needed, see
\cite{Lorenz1969, Leith1978, Basdevant1981, Dalcher1987, Schubert1989, Nicolis1992, Savijarvi1995, Nicolis1995, Trevisan2011}.   

Moreover this description is only valid provided that a short-time linearized description of the error evolution can be performed. 
A very enlightening analysis  on the predictability of turbulent flows based on statistical 
arguments suggests that the propagation of small initial errors in a three-dimensional (3D) flow considerably differs from the one in a 
two-dimensional (2D) flow \cite{Vallis1985,Vallis2006}.  
Specifically, if an error is introduced at small spatial scales in the spectral domain,
the error in a 3D turbulent fluid is predominantly characterized by a nonlinear local cascade propagation that will rapidly
contaminate the largest scales.  For 2D turbulence, the picture is different with an error
dynamics involving both a local nonlinear cascade propagation and the direct amplification of errors along the large spatial scales 
on the same typical timescale. This implies that
a linearized description of the error dynamics at large spatial scales is valid provided that the initial error is small at these large 
scales. 

This feature
can be exploited in the current analysis since the models discussed so far are based on the potential vorticity equation which provides a 2D description 
of the large-scale dynamics of the atmosphere. This result also justifies the use of the Lyapunov exponents for the characterization 
of the error dynamics in such models.

\subsection{Error dynamics in the CS model}

\begin{figure*}
\includegraphics{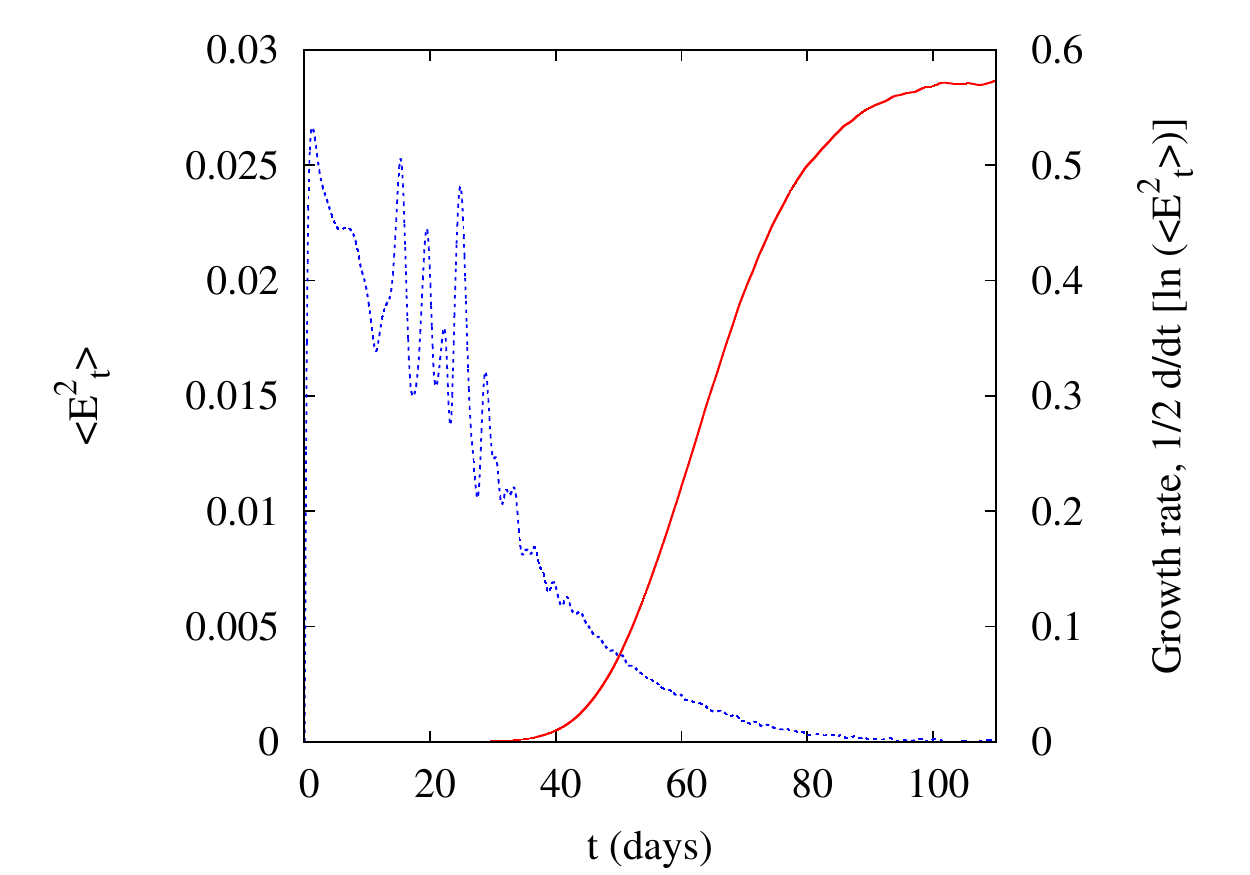}
\caption{
\label{CSerror}
Mean square error (red continuous curve) evolution and error growth rate (blue dotted curve) for the CS model, averaged over 100,000 realizations 
starting from initial conditions sampled on the attractor of the model. 
}
\end{figure*}

Figure \ref{CSerror} displays the error evolution and the growth rate, $1/2 \, d/dt \, ln(<E^2_t>)$,  
as a function of time, as obtained with a very small
initial error perturbing each variable and sampled from a gaussian distribution of mean 0 and variance $10^{-16}$.  The averaging is performed based on 100,000 realizations
starting from different initial conditions on the attractor of the system. The key parameter values used here are 
$\theta_1^*=0.18$ and $n=1.77$.  

The overall behavior is indeed in agreement with the general description presented above 
with an exponential-like phase, followed by a linear regime before the final saturation. 
But when investigating in details the error growth rate (dotted blue curve of Fig. \ref{CSerror}), 
the picture that emerges is very different with a complicate error behavior during
the exponential-like regime. This complicate behavior is characterized by an initial very short error decrease phase for
about 0.08 days 
(not visible on the picture), followed by important variations of the growth rate which reaches a maximum of 0.53 
day$^{-1}$ after 1.34 days. This maximum value is more than two times larger the dominant Lyapunov exponent 
(0.23 day$^{-1}$). This feature -- usually referred to as superexponential -- has been discussed in details 
in \cite{Benzi1989, Nicolis1995, Vannitsem2016} and is mainly associated with the variability of the local Lyapunov exponents.    

\subsection{Error dynamics in the QG3T21 model}

\begin{figure*}
\includegraphics{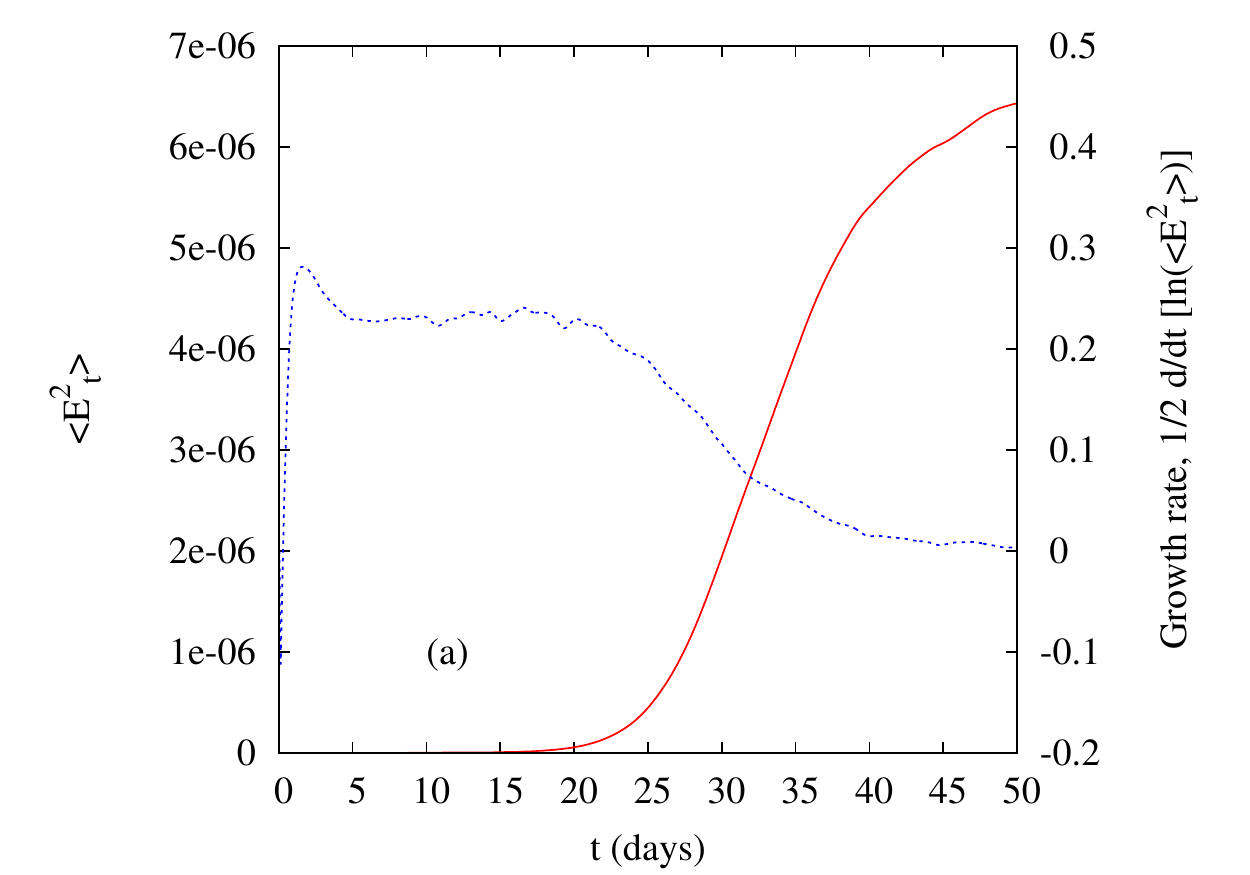}
\includegraphics{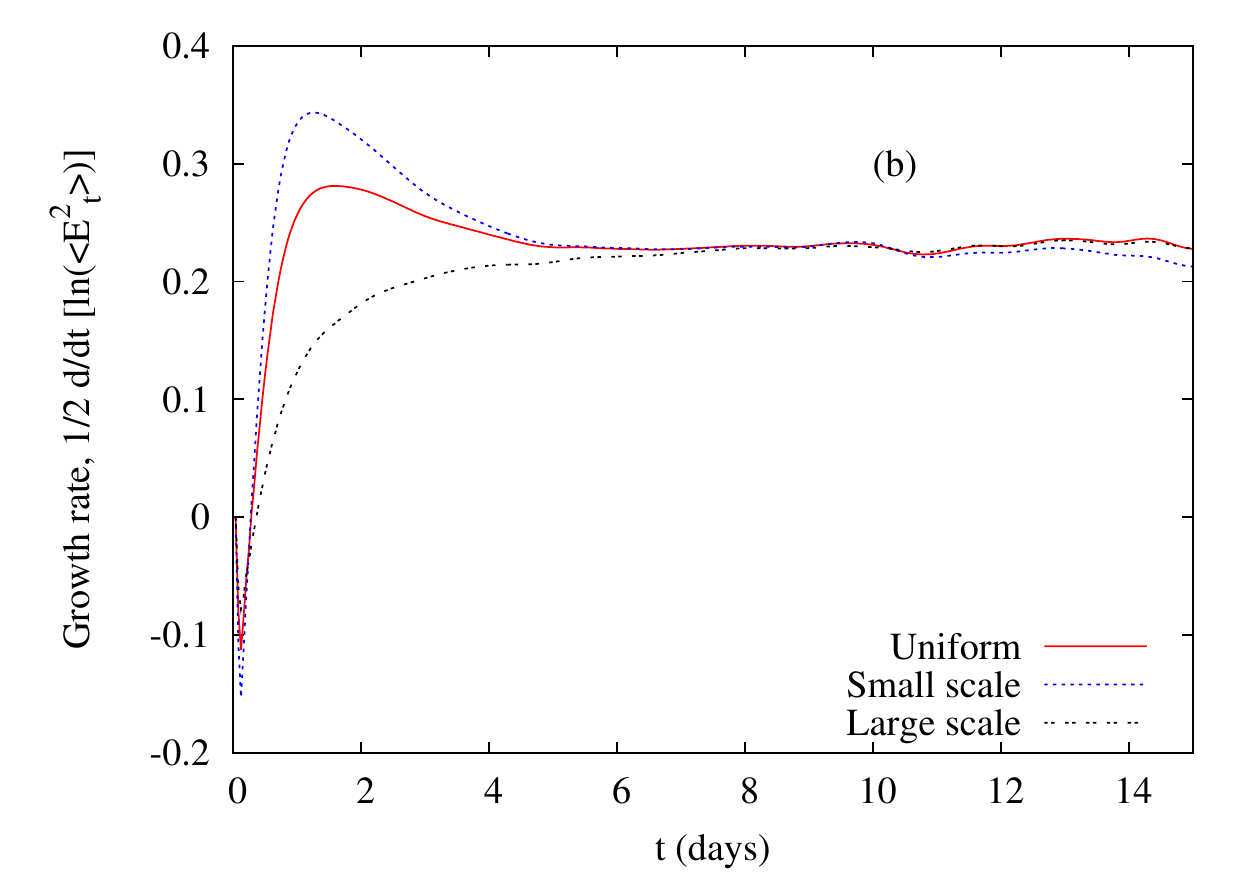}
\caption{
\label{QG3error}
(a) Mean square error evolution (red continuous curve) and growth rate (blue dotted curve) as obtained from 1,000 realizations of the error evolution 
starting from different initial conditions on the attractor of the QGT21L3 model.
(b) Growth rate of the error, (red continuous curve) for a uniform initial error in the spectral domain (in the kinetic energy norm), (black dashed curve) for an initial error predominantly 
located at large spatial scales, and (blue dotted curve) for an initial error predominantly located at small spatial
scales. See also Fig. \ref{QG3errorsp} for the specific repartition of the error in the spectral domain.
}
\end{figure*}

\begin{figure*}
\center
\includegraphics[width=0.49\textwidth]{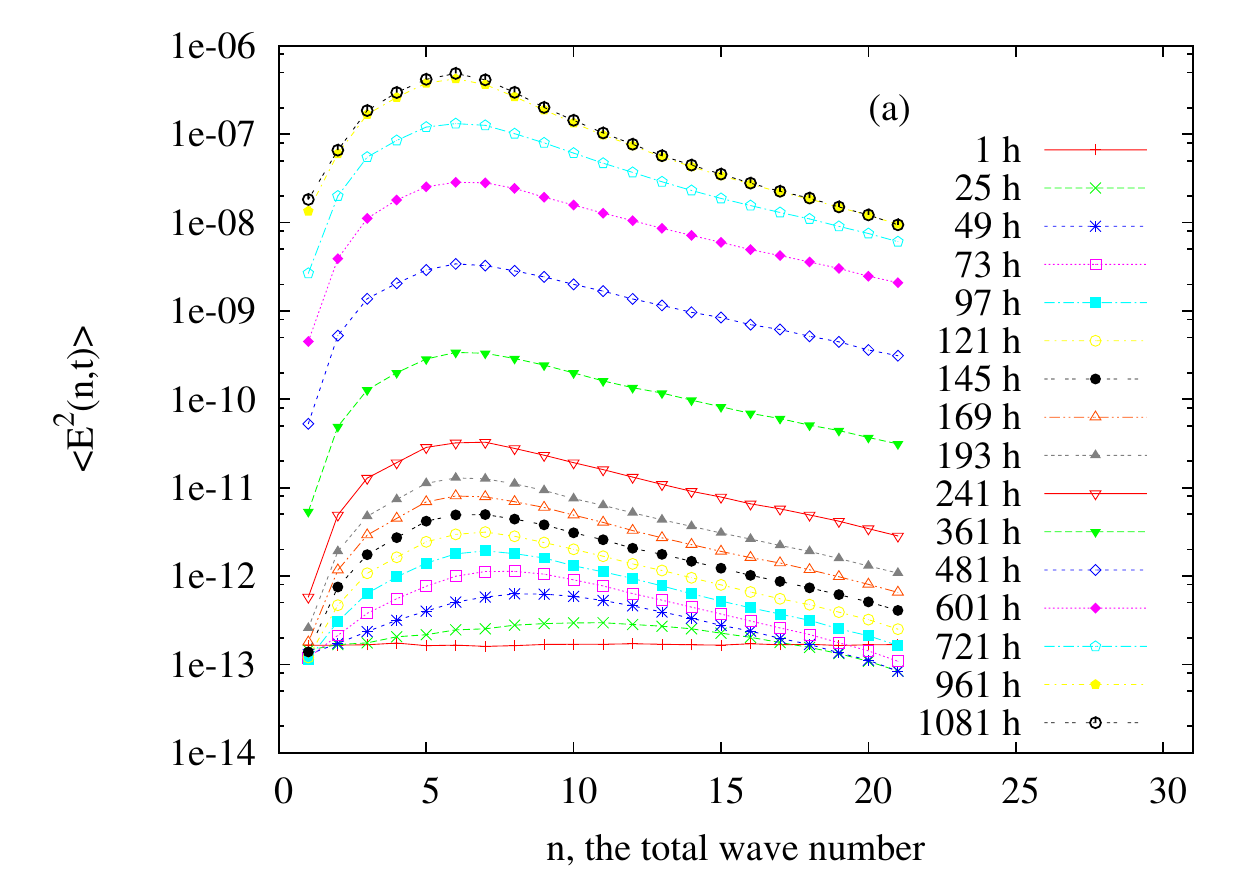}
\includegraphics[width=0.49\textwidth]{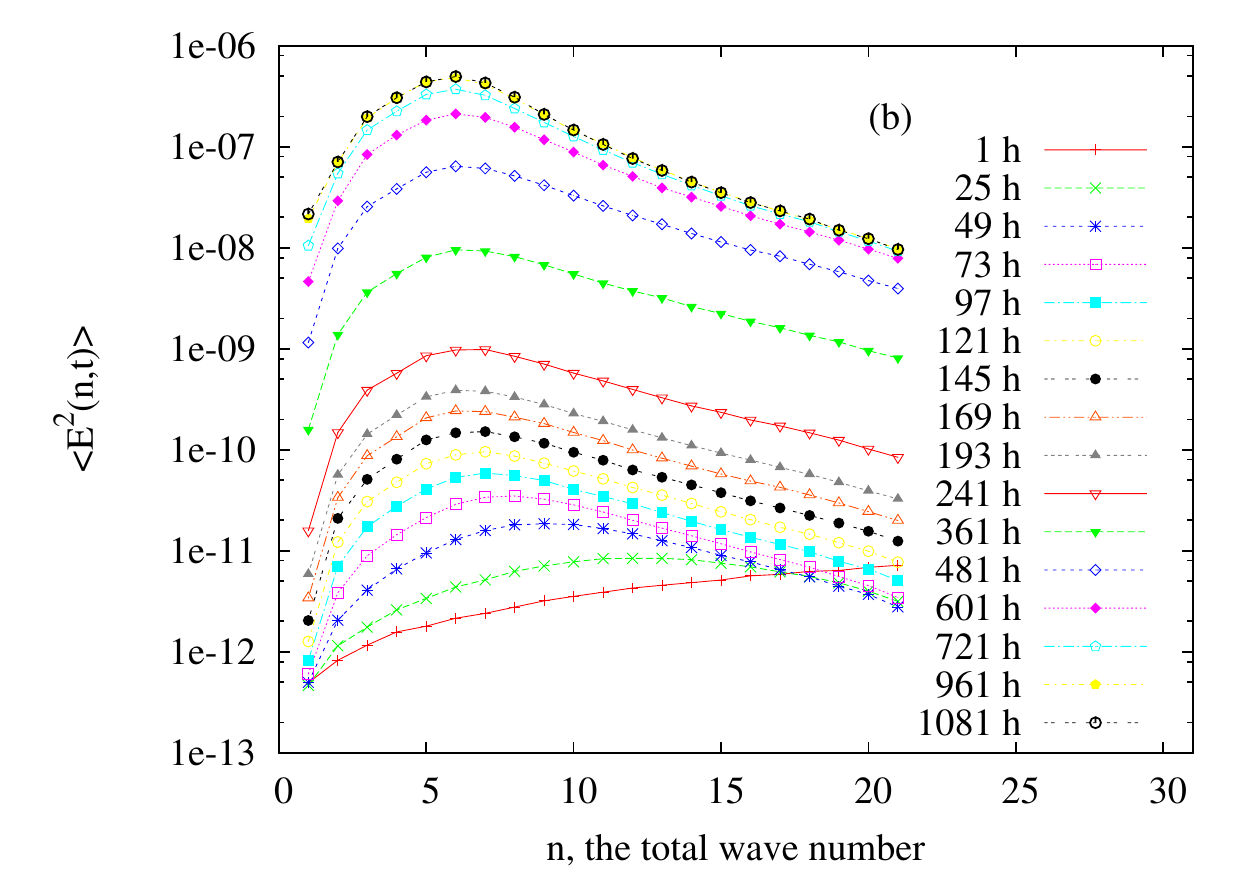}

\includegraphics[width=0.49\textwidth]{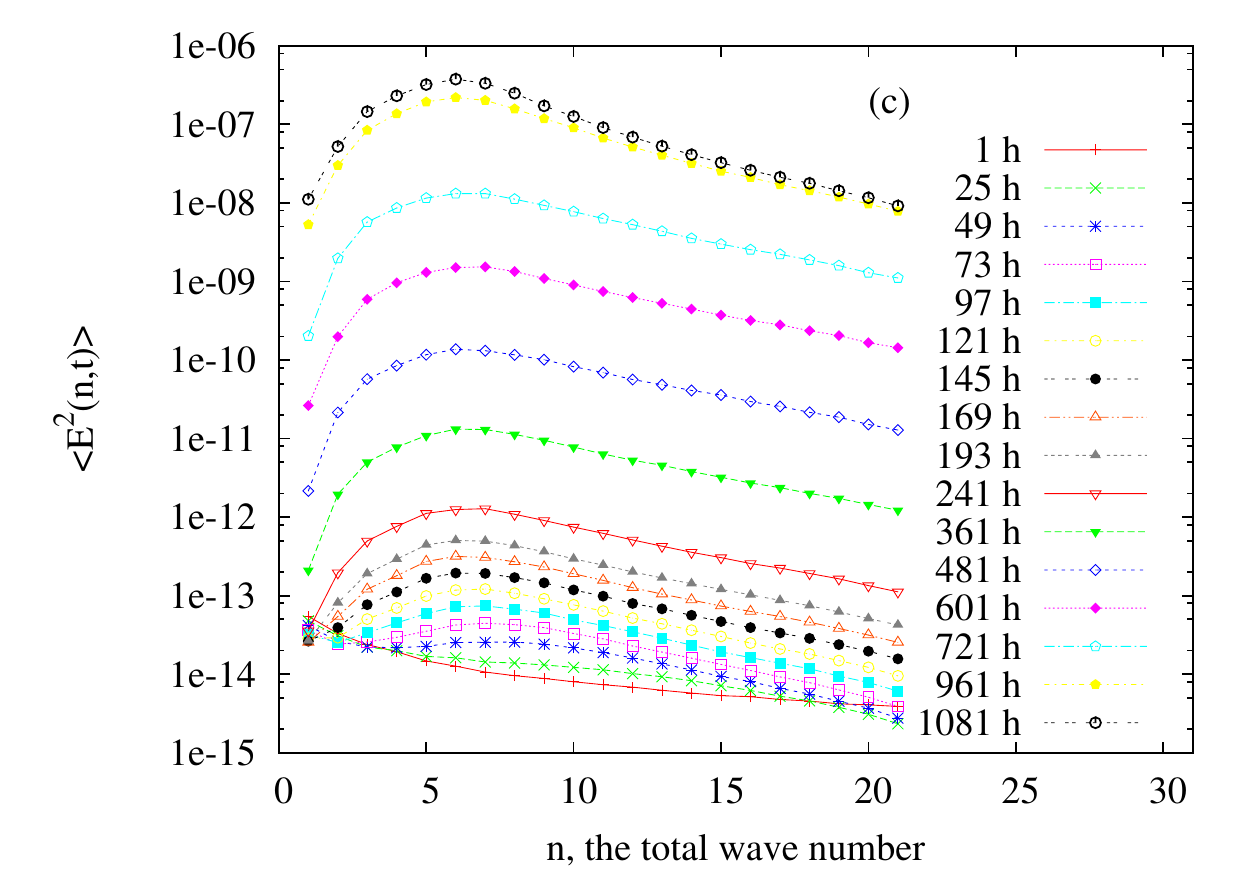}
\caption{
\label{QG3errorsp}
(a) Mean square error evolution for the QGT21L3 model, averaged over 1,000 realizations, as a function of the total wavenumber 
$n$ for different forecasting lead times from 1 hour up to 1081 hours. The initial error is uniformely distributed as a function of $n$ in the energy norm. 
(b) and (c) As in (a) but for an initial error dominating the smallest and largest spatial scales, respectively.}
\end{figure*}

Let us now focus on the error dynamics in the QG3T21 model. Figure \ref{QG3error} displays the mean error evolution (red continuous curve) together
with the error growth rate (blue dotted curve), as obtained from 1,000 realizations starting from different initial conditions on the attractor of the
model. The initial error introduced in the model is a small amplitude error uniformly distributed in the spectral domain as in 
Fig. \ref{QG3errorsp}a. 

A picture similar to the one found in the CS-model can be drawn, with an exponential-like behavior, a linear
amplification of the error and a final saturation phase. The growth rate is however quite different to the one obtained with the CS-model, 
with a maximum not very much larger than the value of the dominant Lyapunov exponent. This difference is the
result of several competing effects: 
(i) the variability of the local Lyapunov exponents associated with the inhomogeneity of the solution's 
attractor, 
(ii) the quasi-continuous  Lyapunov spectrum, 
and (iii) the choice of the initial error (and in particular of its spectral properties). 
The first effect is inducing a super-exponential behavior as already illustrated for the CS model. 
The second one is responsible for the development of a sub-exponential error dynamics as discussed in 
\cite{Nicolis1992b, Vannitsem1996}. 
The third one is modulating the initial decrease of the error and the (nonlinear) transfer of errors across scales.       

In the present model, the third effect plays a dominant role as illustrated in the comparison of growth rates for different 
random initial errors in Fig. \ref{QG3error}b. In this case when errors are introduced at large spatial scales (black dashed curve), the growth rate of the error 
is slowly increasing and reaches the value of the dominant Lyapunov exponent after about 10 days. When errors are introduced at
small spatial scales (blue dotted curve), the growth rate increases rapidly and reaches a maximum after about 1.5 days which is larger than the
one obtained with a uniform initial error (red continuous curve).     

This  dynamics is better analyzed in the spectral domain. Figure \ref{QG3errorsp} shows the error evolution as a function
of the total wave number $n$ at 500 hPa. The norm used to evaluate the error is the kinetic energy norm. 
A first interesting feature is the decrease of
the error at very large and very small spatial scales, while the error increases at intermediate scales. As stated in \cite{Vannitsem1997}, 
these intermediate scales are the ones at which the dominant (Backward) Lyapunov vectors are operating, while the stable (Backward) 
Lyapunov vectors are mostly acting at very large and small spatial scales, inducing the specific error behavior observed in this figure.  

When the initial error is confined to the very small spatial scales, a transfer (through nonlinear interactions) is occurring toward larger scales,
together with the amplification along the dominant (backward) Lyapunov vectors, inducing a larger amplification rate of the error as
illustrated by the blue dotted curve in Fig. \ref{QG3error}b. When the intial error is essentially confined at large spatial scales the error evolution 
essentially displays an amplification according to the Lyapunov instabilities, without important transfer toward smaller scales. These behaviors are
illustrated in Figs \ref{QG3errorsp}b-c.

The error dynamics is considerably dependent on the specific scale at which it is introduced as illustrated in Fig. \ref{QG3errorsp}, with a faster
growth when located at small spatial scales. This feature has been exploited in the development of probabilistic forecasts in the 90th, for which strong
growth were search for in order to get a sufficient variability in the multiple integrations of the ensemble forecasting systems 
made by the meteorological centers, e.g. \cite{Molteni1993, Buizza1995, Ehrendorfer1995, Szunyogh1997, Gelaro2002}.   

\subsection{Error dynamics in the ocean-atmosphere model}

Up to now the error dynamics has been discussed for a system -- the atmosphere --  displaying a variability {\sv on a range of} timescales relatively 
close to each other, i.e. typically from a few hours up to a few days. When this system is coupled to an ocean, highly different 
timescales are
involved as already illustrated in the Lyapunov spectra of Section \ref{OA-lyap}. The question is therefore to know what is the nature of the error dynamics
in these different sub-systems and what is the impact of {\sv the ocean-atmosphere coupling on }predictability.    
As in Section \ref{OA-lyap} we will focus on the two parameter sets explored, with and without low-frequency variability.

\begin{figure*}
\includegraphics{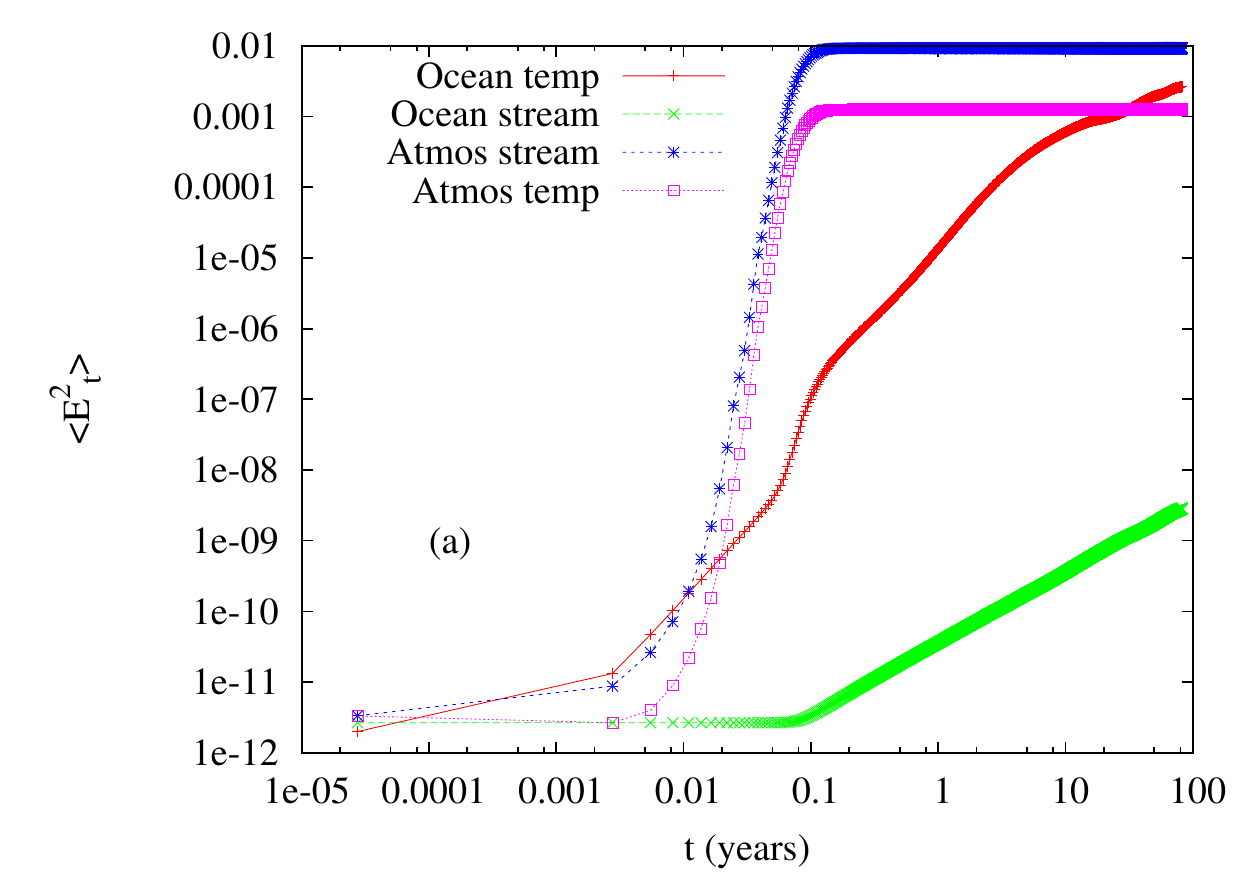}
\includegraphics{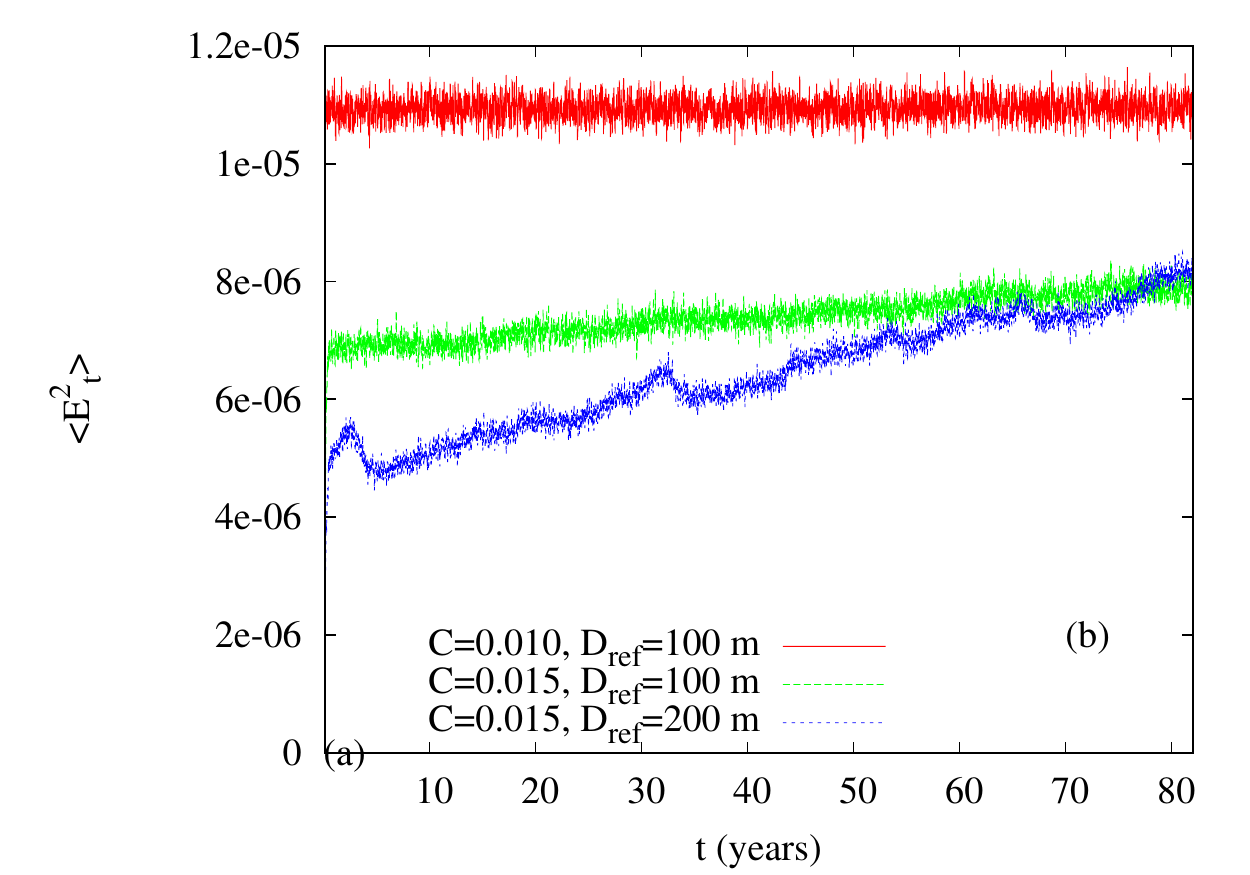}
\caption{ 
\label{AOerror}
(a) Mean square error evolution for the {\svn ocean-atmosphere model}, for the atmospheric
barotropic streamfunction (blue stars), the atmospheric temperature (magenta open squares), the ocean streamfunction (green crosses), and 
the ocean temperature (red pluses).  The model parameter values used are $S_o=310$ W m$^{-2}$, $D_{ref}=100.$ m, $\kappa=1$ and with $C=0.010$ kg m$^{-2}$ s$^{-1}$. 
(b) as in (a) but for the first barotropic streamfunction mode, $\psi_1$, only and for parameters: $S_o=310$ W m$^{-2}$, $D_{ref}=100$ m, $\kappa=1$,         
$C=0.010$ kg m$^{-2}$ s$^{-1}$ already used in Fig. (\ref{AOerror}a) (red continuous curve), $S_o=310$ W m$^{-2}$, $D_{ref}=100$ m, $\kappa=1$, 
$C=0.015$ kg m$^{-2}$ s$^{-1}$ (green dashed curve), and $S_o=310$ W m$^{-2}$, $D_{ref}=200$ m, $\kappa=1$, $C=0.015$ kg m$^{-2}$ s$^{-1}$ (blue dotted curve). 
 }
\end{figure*}

Figure \ref{AOerror}a displays the mean square error evolution for the different fields present in the coupled A-O model, namely 
the atmospheric 
barotropic streamfunction (blue line referred as "Atmos stream"), the atmospheric temperature (magenta line with open squares referred as "Atmos temp"),
the ocean streamfunction (green curve with crosses referred as "Ocean stream"), and the ocean temperature (red curve with pluses referred as "Ocean temp").
{\sv These curves are obtained with the parameter values} corresponding
to the red attractor of Fig. \ref{OAattractor}.
The error rapidly amplifies for both atmospheric fields and saturates at a constant value after about 1 month (1/12 years). For the ocean
the picture is very different with an increase of the error that persists beyond 100 years revealing a high potential of predictability.

If now one considers cases for which an attractor is developing around an unstable periodic orbit as the green attractor of Fig. \ref{OAattractor}, the picture
is very different as illustrated in Figure \ref{AOerror}b displaying the mean square error evolution of the first
barotropic mode, $\psi_1$, of the atmosphere.    
{\sv The two lower curves} associated with the error dynamics when the attractors are developing around an unstable periodic orbit 
(typically the green attractor of Fig. \ref{OAattractor}) are still increasing beyond the 30 days limit detected in 
Fig. \ref{AOerror}a.
The behavior of the mean square error is in this case linear as a function of time suggesting a diffusive dynamics. If now we assume that
the mean square error evolution can be modelled as $D (t-t_0)$ for $t > t_0=30$ days, the diffusion parameter,  $D$, is  
larger when the reference depth of the ocean $D_{ref}$ is larger.   

This finding suggests that when the attractor is developing around the long unstable periodic orbit arising from the coupling between the ocean 
and the atmosphere \cite{Vannitsem2015b, Vannitsem2015a}, some atmospheric modes can be predictable for periods much longer than the timescale
of a few days (or weeks) typical of the atmospheric dynamics. This feature provides some hope in performing long term forecasts of specific observables in
the atmosphere at seasonal, interannual and decadal timescales.   

\section{Conclusions and perspectives}

The Lyapunov instability properties and the error dynamics in large scale flows of the atmosphere, coupled or uncoupled to 
the ocean, have been explored in a hierarchy of models from low-order, $ O(10)$ variables, up to intermediate order, $ O(1,000)$. Two major trends
are emerging when the number of variables is increased: (i) the Lyapunov spectrum can have a number of positive exponents of the order of $O(100)$ (in the standard version of the QG3T21 intermediate-order model), implying
that the attractor dimension of the modelled atmosphere is high dimensional, and (ii) the variability of the local instability 
properties associated with the largest Lyapunov exponent decreases. 

The first trend is in line with the mathematical findings on the bounds of the attractor dimension, indicating that the atmosphere 
is living on a finite but high dimensional attractor \cite{Wang1992, Lions1997}. 
{\sv The second one is suggesting a smoothing of the inhomogeneity of the attractor when the number of variables is increased, and it
opens important questions on the  characteristics of the instability properties of the solutions in the limit of an infinite number of modes, i.e. for the continuous partial differential 
equations (quasi-geostrophic equations) discussed in Section \ref{model}.}  

From a practical point of view, it is well known that the predictability of large-scale weather patterns is highly dependent on the specific initial weather
situation selected \cite{Ehrendorfer1997, Yoden2007}. This suggests that this inhomogeneity is also present in high resolution atmospheric models but to which
extent it resembles the one described in the quasi-geostrophic model or not, is still open. A possible way to address this
question is to extend the Lyapunov analysis to high-dimensional quasi-geostrophic models of the order of $O(10,000, 100,000)$ variables.   

Yet, even if this analysis would provide interesting information on the smoothness of {\sv the attractor's instability properties}, it is not the end of the story. When the resolution
of such models is increased, their validity in describing the dynamics of the atmosphere (and the ocean) is questionable. In this case one must go back
to the original primitive equations discussed in the beginning of Section \ref{model}, and to study the impact of
the dynamics of the additional variables.  

In this perspective, one particularly interesting work has been done by Uboldi and Trevisan \cite{Uboldi2015} 
in which they have studied the
full primitive equations integrated at the cloud-resolving scale of 2.2 km resolution, with 50 vertical levels. In this
context they have studied the Bred vectors {\sv that are {\it finite-size} unstable structures emerging along the model trajectory}. 
If their amplitudes is sufficiently small these are providing information on 
instability properties of the flow  closely related to the Backward Lyapunov vectors \cite{Szunyogh1997}.  
In the case of small-amplitude Bred vectors, they found that the dominant ones describe small convective-scale instabilities
and that their number is very large. Moreover their growth rates are much larger 
than any perturbation acting at the large spatial scales of the flow (and much larger than the ones displayed in the present review). 
This result is consistent with the 
theoretical considerations presented by Vallis \cite{Vallis1985, Vallis2006}, indicating that the dynamics of the error
in a three-dimensional turbulent system -- as it is the case in the high-resolution experiment of Uboldi and Trevisan --
is saturating rapidly at all the scales below 10 km, and that predictability is limited to a few hours. 

These results are obviously suggesting that the development of very high resolution models at 1 km or smaller
would lead to a very small gain in terms of predictability due to the rapid saturation of the error at small spatial scales, 
and one should therefore tempered us in developing such models except if the forecast at a lead time of one or two hours is 
providing important information for the society. In such a case a very performant data assimilation system would also be
needed in order to reduce considerably the actual level of error. This is a very expensive process  
in terms of maintenance (man power), high-quality observing systems and computer power. 
In the sake of reducing the potential cost,
it might be better to avoid such high resolution forecasts.
Yet, these very high resolution forecasts should  
be necessary in some very specific and dangerous weather situations arising from time to time. {\sv In an operational environment 
a procedure should be developed in order} to evaluate the necessity to perform very high resolution forecasts. {\sv This
procedure could be based on storminess warnings that can be gathered from lower resolution model integrations}.

Nowadays operational forecasting systems contain stochastic schemes emulating the presence of model errors.
This approach allows for getting more reliable ensemble forecasts as discussed in \cite{Berner2009, Buizza2010}, but also for improving
climatological aspects of the models such as correcting the mean or the variance of specific variables \cite[e.g.][]{Weisheimer2014}.
This however implies that an increase of uncertainty is introduced, inducing in particular a larger error for short times 
\cite{Nicolis2003, Nicolis2004}. 
Another important issue is therefore to understand the role of stochastic forcings on the predictability of atmospheric and oceanic 
flows. Theoretical and practical analyses are therefore necessary in the line of the works of 
\cite{Seki1993, Chu2005, Vannitsem2014b, Ghil2017}. 

Finally the problem of forecasting atmospheric phenomena on timescales longer than the typical limit of weather forecasts, say 10 days,
is a challenging problem for our society. As we demonstrate in the present review in the context of an idealized coupled ocean-atmosphere
model, long term forecasts of specific atmospheric variables are possible, provided the atmosphere is (strongly) coupled to climate
components with longer typical timescales. This increase of forecast skill at seasonal and interannual timescales 
is already well known in the Tropical regions due to the strong coupling between the ocean and the atmosphere \cite[e.g.][]{Phil1990}. 
Several climate forecasting models are also currently suggesting that there is some skill of the climate
system at midlatitudes at seasonal and interannual timescales, although the signal is rather weak \cite[e.g.][]{Smith2016}.  
The understanding of the origin of this long term skill and our ability to improve the long term forecasts are still fields of research in their infancy. 
The exploration of the coupling bewteen the ocean and the atmosphere (and other climate components) in the context of low and intermediate
order models are important steps in that direction \cite{Kravtsov2008, Brachet2012, Vannitsem2015a, Vannitsem2015b}.  

\bigskip

\begin{acknowledgments}
The comments of C. Nicolis and two anonymous reviewers on an earlier version of this manuscript were highly appreciated.
This work is supported, in part, by the Belgian Science Policy Office under contract BR/121/A2/STOCHCLIM.
\end{acknowledgments}

{}


\end{document}